\documentclass[final]{siamltex}

\usepackage{epsfig, amsmath, amssymb}

\newcommand{\keyw}[1]{{\bf #1}}

\title{A fast multigrid algorithm for energy minimization under planar
density constraints}

\author{Dorit Ron \thanks{The Weizmann Institute of Science, dorit.ron@weizmann.ac.il} \and Ilya Safro
\thanks{Argonne National Laboratory, safro@mcs.anl.gov} \and  Achi Brandt
\thanks{The Weizmann Institute of Science} }

\begin{document}

\maketitle
\begin{abstract}
The two-dimensional layout optimization problem reinforced by the
efficient space utilization demand has a wide spectrum of
practical applications. Formulating the problem as a nonlinear
minimization problem under planar equality and/or inequality
density constraints, we present a linear time multigrid algorithm
for solving correction to this problem. The method is demonstrated
on various graph drawing (visualization) instances.

\end{abstract}


\pagestyle{myheadings}
\thispagestyle{plain}
\markboth{D. Ron AND I. Safro AND A. Brandt}{A multigrid approach for constrained optimization}

\bibliographystyle{plain}



\newcommand{\solution}{\par\noindent{\bf Solution:}\quad}
\newcommand{\minla}{Minimum Linear Arrangement problem }

\newcommand{\genr}{\mathfrak{E}}
\newcommand{\geqd}{\mathfrak{eqd}}

\newcommand{\eqd}{\mathfrak{eqd}}
\newcommand{\gr}{\mathcal{G}}
\newcommand{\nar}{\mathfrak{a}}
\newcommand{\wnd}{\mathcal{W}}
\newcommand{\pgr}{\mathcal{P}(\mathcal{G})}
\newcommand{\sgr}{\mathcal{S}(\mathcal{G})}
\newcommand{\bgr}{\mathcal{B}(\mathcal{G})}
\newcommand{\bwn}{\overline{\mathcal{B}}(\mathcal{W})}
\newcommand{\bpwn}{\mathcal{B}(\mathcal{W})}
\newcommand{\bugr}{\mathcal{B}_u(\mathcal{G})}
\newcommand{\bvgr}{\mathcal{B}_v(\mathcal{G})}
\newcommand{\ar}{\mathcal{A}}
\newcommand{\ineigh}{\mathcal{P}_i}
\newcommand{\jneigh}{\mathcal{P}_j}
\newcommand{\sar}{\Upsilon}
\newcommand{\mar}{\upsilon}
\newcommand{\mdl}[1]{\text{(mod }#1\text{)}}
\newcommand{\Lag}{\mathfrak{L}}
\newcommand{\bu}{{\bf u}}
\newcommand{\bU}{{\bf U}}
\newcommand{\bpgr}{\bold{\mathcal{P}}_{\bu}}
\newcommand{\bpgrw}{\bold{\mathcal{P}}_{\bu_{\wnd}}}
\newcommand{\bpw}{\bold{\mathcal{P}}_{\wnd}}


\section{Introduction}\label{sIntro}
\par
\begin{figure}[h]\label{snake}
\vbox{\center\includegraphics[width=10cm]{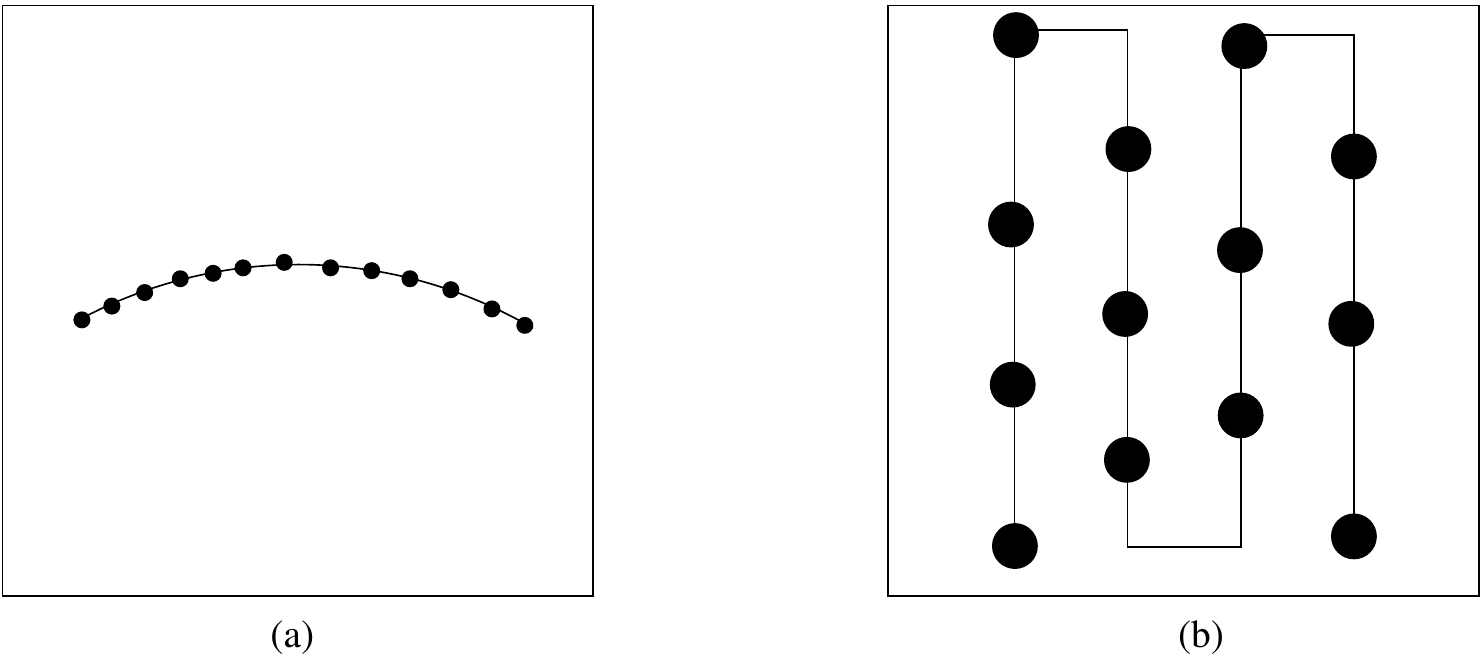}}
\caption{Possible ways to draw the "snake"-like graph: (a) when
the drawing area is not used, the size of the nodes must
decrease; and (b) a clearer picture is obtained when the space is
used efficiently. }
\end{figure}

 The optimization problem addressed
in this paper is to find an optimal layout of a set of
two-dimensional objects such that (a) the total length of the
given connections between these objects will be minimal, (b) the
overlapping between objects will be as little as possible, and, (c)
the two-dimensional space will be well used. This class of
problems can be modelled by a graph in which every vertex has a
predefined shape and area and each edge has a predefined weight.
While the first two conditions are straightforward, the third
requirement can be made concrete in different ways. To see its usefulness, consider for example, the problem of
drawing the "snake"-like graph shown in Figure \ref{snake}(a).
Most graph drawing algorithms would draw it as a line or a chord.
In that case, when the number of nodes is big, the space is
used very inefficiently, and the size of the nodes must
decrease.
One possible
efficient space utilization for the graph "snake" is presented in
Figure \ref{snake}(b).
\par In many theoretical and industrial fields, this class of problems
is often addressed and actually poses a computational bottleneck.
In this work we present a multilevel solver for a model that
describes the core part of those applications, namely, the problem
of minimizing a quadratic energy functional under planar
constraints that bound the allowed amount of material (total
areas of objects) in various subdomains of the entire domain under
consideration.
\par Given an initial arrangement, the main contribution of this
work is to enable a {\it fast} rearrangement of the entities under
consideration into a more evenly distributed state over the entire
defined domain. This process is done by introducing a sequence of
finer and finer grids over the domain and demanding at each scale
{\it equidensity}, that is, meeting equality or inequality
constraints at each grid square, stating how much material it may
(at most) contain. Since many variables are involved and since the
needed updates may be large, we introduce a new set of {\it
displacement} variables attached to the introduced grid points,
which enables {\it collective} moves of many original variables at
a time, at different scales including large displacements. The use
of such multiscale moves has two main purposes: to enable processing in various scales and to {\it efficiently} solve the (large) system of equations of energy
minimization under equidensity demands. The system of equations of
the finer scales, when many unknowns are involved, is solved by a
combination of well-known multigrid techniques (see
\cite{Brandt:1977:MLAa,vlsicad,mgbooktrott}), namely, the {\it
Correction Scheme} for the energy minimization part and the {\it
Full Approximation Scheme} for the inequality equidensity
constraints defined over the grid's squares. We assume here that
the minimization energy functional has a quadratic form, but other
functionals can be used via quadratization. The entire algorithm
solves the nonlinear minimization problem by applying successive
steps of corrections, each using a linearized system of
equations.

\par Clearly, for each specific
application, one has to tune the general algorithm to respect the
particular task at hand.
We have chosen here to demonstrate the performance of our solver
on some instances of the graph visualization problem showing the
efficient use of the given domain. Let us review a few
applications that have motivated our research.
\par {\bf Graph visualization} addresses the problem of constructing a geometric representation of graphs
and has important applications to many technologies. There are
many different demands for graph visualization problems, such as
draw a graph with a minimum number of edge crossings, or a minimum
total edge length, or a predefined angular resolution (for a
complete survey, see \cite{gd-book}). The ability to achieve a
compact picture (without overlapping) is of great importance,
since area-efficient drawings are essential in practical
visualization applications where screen space is one of the most
valuable commodities. One of the most popular strategies that does
address these questions is the force directed method
\cite{eades1984} which has a quadratic running time if all
pairwise vertex forces are taken into account. There are several
successful multilevel algorithms \cite{DBLP:conf/gd/2000}
developed to improve the method's complexity. However, reducing
the running time in these models usually means a loss of
information regarding those forces.
\par {\bf Representation of higraphs}. Higraphs, a combination and extension of graphs and Euler/Venn diagrams,
were defined by Harel in \cite{harel88}. Higraphs extend the basic
structure of graphs and hypergraphs to allow vertices to describe
inclusion relationships. Adjacency of such vertices is used to
denote set-theoretic Cartesian products. Higraphs have been shown
to be useful for the expression of many different semantics and
underlie many visual languages, such as statecharts and object
model diagrams. The well-known force-directed method has been
extended to enable handling the visualization of higraphs
\cite{harelinger}. For small higraphs it has indeed yielded
nice results; but because of its high complexity, it poses efficiency
challenges when used for larger higraphs.
\par {\bf Facility location problem}. In this class of problems the goal is to locate a number
of facilities within a minimized distance from the clients. In many industrial versions of the problem
there exist additional demands such as the minimization of the routing between the facilities and various
space constraints (e.g., the factory planning problem) while given a total area on which the facilities and
clients could be located (for a complete survey, see \cite{dreznerfacility}).
\par {\bf Wireless networks and coverage problems} have a broad range of applications in the military,
surveillance, environment monitoring, and healthcare fields. In
these problems, having a limited number of resources (like antenna
or sensor), one has to cover the area on which many demand points
are distributed and have to be serviced. In many practical
applications there are predefined connections between these
resources that can be modeled as a graph
\cite{meguerdichian01coverage, cardei-energyefficient,
citeulike:717044}.
\par {\bf The placement problem}. The electronics industry has achieved a phenomenal growth over the past
two decades, mainly due to the rapid advances in integration technologies and large-scale systems design - in short,
due to the advent of VLSI. The number of applications of integrated circuits in high-performance computing,
telecommunications, and consumer electronics has been rising steadily, and at a very fast pace. Typically,
the required computational power of these applications is the driving force for the fast development of this field.
The global placement is one of the most challenging problems during VLSI layout synthesis. In this application
the modules must be placed in such a way that the chip can be processed at the detailed placement stage and
then routed efficiently under many different constraints. This should be accomplished in a reasonable computation
time even for circuits with millions of modules since it is one of the bottlenecks of the design process.
For a most recent survey of the placement techniques see \cite{vlsi2007book}.
\par The paper is organized as follows. The problem definition is described in Section \ref{prob-def}.
The multilevel formulation and solver are presented in Section
\ref{ml-formulation}. Examples of graph drawing layout corrections
are demonstrated in Section \ref{sExamples}.

\section{Problem definition}\label{prob-def}
\par
\begin{figure}[h]
\vbox{\center\includegraphics[width=6.5cm]{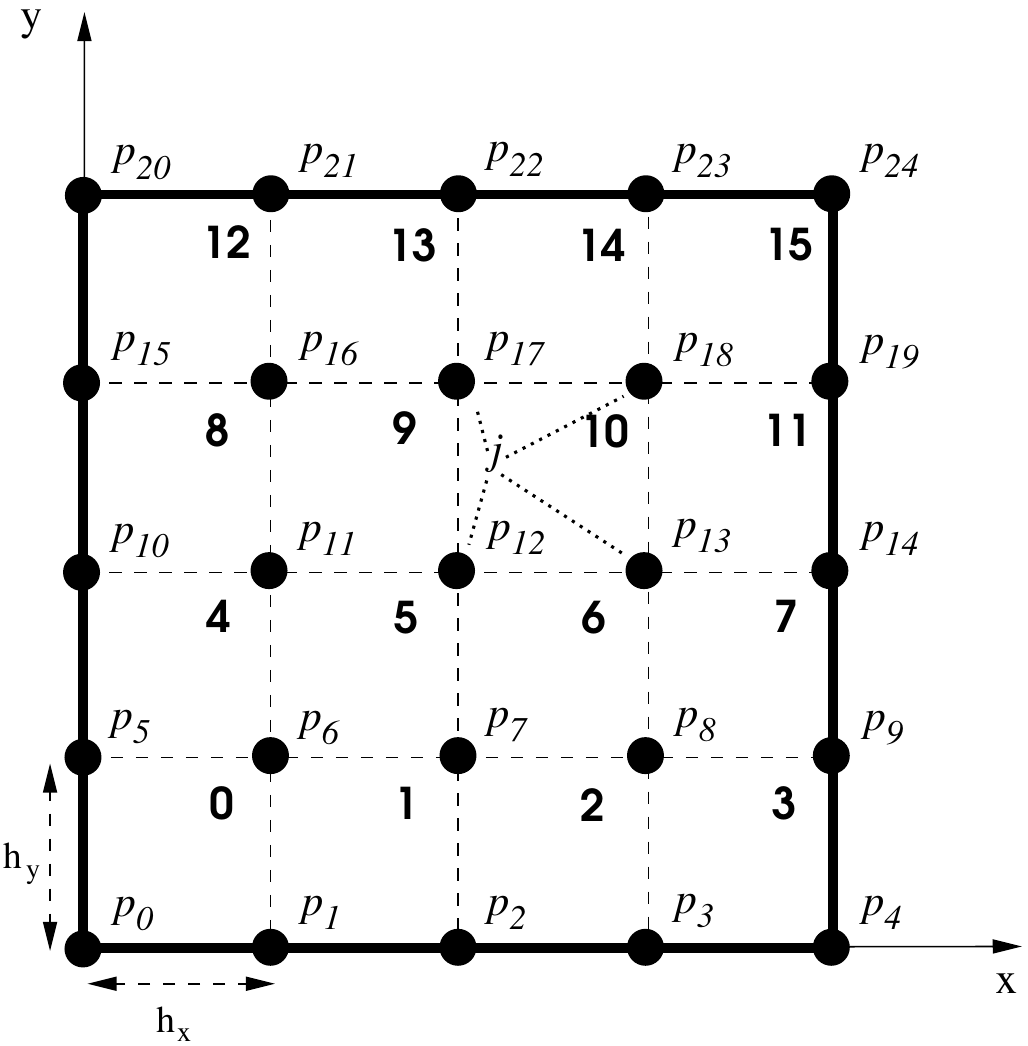}}
\caption{Example of a grid $\gr$ with $25$ grid points and $16$
squares.
The grid points and squares are labeled by $p_i$ and bold numbers,
respectively.}\label{chipgrid}
\end{figure}

\begin{figure}[h]
\vbox{\center\includegraphics[width=5cm]{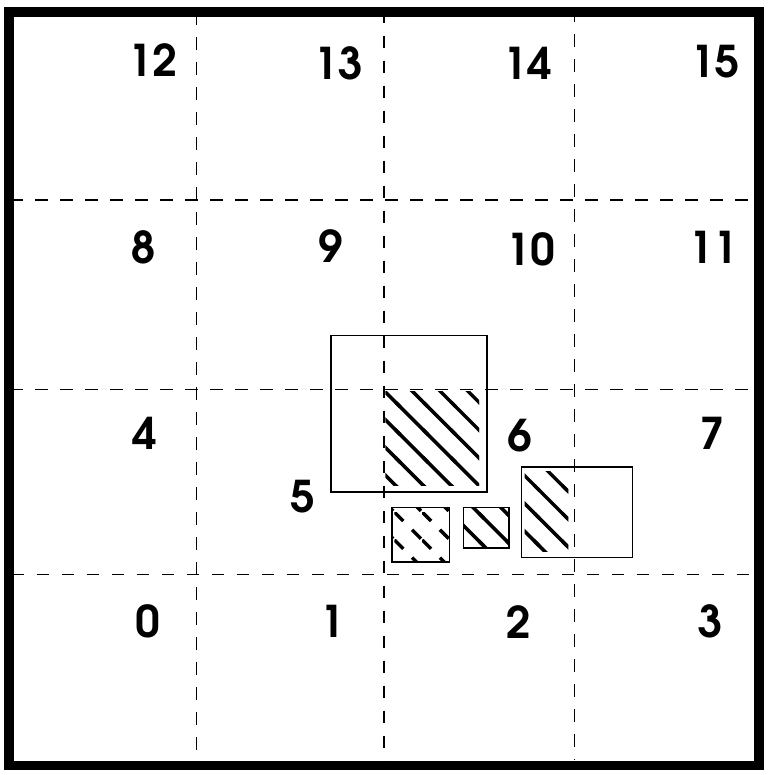}}
\caption{Example of $\sar(s)$ for square 6. The total area of vertices
overlapping with square 6 is dashed.} \label{partarea2}
\end{figure}

 Given a weighted undirected graph $G=(V, E)$, let $v(i)>0$ be the (rectangular)
area of vertex (node) $i\in V$, $i=1,...,|V|$,
and $w_{ij}$ the non-negative weight of the edge $ij$ between
nodes $i$ and $j$ ($w_{ij}=0$ if $ij\notin E$). Also, assume a twodimensional {\it initial} layout is given; that is, the center of
mass of node $i$ is considered to be located at $(\tilde{x}_i,
\tilde{y}_i)$ within a given rectangular domain. The purpose of
the optimization problem we consider is to modify the initial
assignment $(\tilde{x}, \tilde{y})$ by $(\delta_{x}, \delta_{y})$
so as to minimize the quadratic functional
\begin{equation}\label{first-probform}
\mathfrak{E}(\delta_{x}, \delta_{y}) = \frac{1}{2}\sum_{ij\in E}
w_{ij}\big(
(\tilde{x}_i+\delta_{x_i}-\tilde{x}_j-\delta_{x_j})^2+(\tilde{y}_i+\delta_{y_i}-\tilde{y}_j-\delta_{y_j})^2\big)
~,
\end{equation}
subject to some {\it equidensity} demands on the area distribution
of the nodes within the given rectangle. To apply such
constraints, we discretize the domain by a standard grid $\gr$
consisting of a set of squares $\sgr$, where each square $s\in
\sgr$ is of area $\ar = h_x h_y$ and $h_x$ and $h_y$ are the mesh
sizes of $\gr$ in the $x$ and $y$ directions, respectively (see
Figure \ref{chipgrid}). Denote by $\sar(s)$ the total area of the
vertices overlapping with the square $s$; that is, $\sar(s)$ is
the sum over all the nodes coinsiding with $s$, each contributing
the (possibly partial) area that overlaps with $s$ (see Figure
\ref{partarea2}).

\par The {\it planar} constraints (i.e., the constraints that are distributed over the 2D plane,
where each constraint defines a demand regarding some bounded
area) can then simply state how much area is required to be in
every square; that is, for each square $s\in \sgr$ the constraint
is either $\sar(s)=M(s)$, or $\sar(s)\leq M(s)$, where $M(s)$ is
the amount of nodes area desired or allowed for square $s$.
\par The constrained optimization problem with equality or inequality formulation can thus be summarized by the
following
\begin{equation}\label{full-problem}
\begin{array}{ll}
\mbox{\textbf{minimize}}  & \mathfrak{E} ~~~ (\text{given by } (\ref{first-probform})) \\
\mbox{\textbf{subject to}} & \sar(s)=(\leq) M(s)  ~,~ \forall s
\in \sgr.
\end{array}
\end{equation}

\section{The multilevel formulation and solver}\label{ml-formulation}
\par The aim of the current work is to provide a {\it fast} first-order
correction to the given approximate solution; that is, we are
looking for such a displacement that would in some optimal sense
(to be defined below) improve the planar equidensity demands
and/or decrease $\mathfrak{E}$. (Note that unconstrained
minimization of $\mathfrak{E}$ will bring all nodes to overlap at
a single point, and thus we may often observe an {\it increase} in
$\mathfrak{E}$ upon removing some of the initial overlap.)
\par To enable a direct use of the multigrid paradigm, and motivated by the
need to perform {\it collective} moves of nodes (as explained in
the introduction), we have actually reformulated the problem
(\ref{first-probform}) as described in Section \ref{sProbRef}. The
multilevel solver of the (reformulated) system
(\ref{full-ineq-problem}) below is introduced in Section
\ref{sMultSolver2}. This system of equations actually has to be
solved for a {\it sequence} of different grid sizes to enhance the
overall equidensity for a variety of scales as presented in
Section \ref{sExternalCycle}.

\subsection{Formulation of the correction problem}\label{sProbRef}
\par We have first introduced two new sets of variables $u$ and $v$ that correspond to {\it displacements}
 in the horizontal
 and vertical directions, respectively. These variables are located at the {\it grid points} $\pgr$
 which are sequentially counted from $0$ to $|\pgr|-1$ as shown in Figure \ref{chipgrid}.
 Each point $p\in \pgr$ is associated with two variables $u_p$ and $v_p$ that influence the
 displacements
 of all the nodes located in the (up to four) squares intersecting at $p$.
 For example, the horizontal update of (the center of mass of) node $j$,
 depicted in Figure \ref{chipgrid}, is obtained
 from points $p_{12}, p_{13}, p_{17}$ and $p_{18}$:
\[
x_j\leftarrow
x_j+\alpha_{12,j}u_{12}+\alpha_{13,j}u_{13}+\alpha_{17,j}u_{17}+\alpha_{18,j}u_{18},
\]
where $\alpha_{12},\alpha_{13},\alpha_{17}$ and $\alpha_{18}$, are
the standard bilinear interpolation coefficients (their sum equals
1). The vertical coordinate $y_j$ is updated from the $v$
variables using the {\it same} coefficients.
\par For a node $i$ denote the
set of four closest points in $\pgr$ (the corners of the square
its center of mass is located at) by $c(i)$.
The new quadratic energy functional we
would like to minimize for $u$ and $v$ given a current layout
$(\tilde{x},\tilde{y})$ of $G$ (i.e., the coordinates of node $i$
are initialized with $(\tilde{x}_i,\tilde{y}_i))$ is
\begin{multline}\label{min-func}
\mathfrak{E}(u, v) = \\
\frac{1}{2}\sum_{ij\in E} w_{ij} \biggl[ \biggl( \tilde{x}_i +
\sum_{p\in c(i)}\alpha_{pi} u_p - \tilde{x}_j - \sum_{p\in
c(j)}\alpha_{pj} u_p\biggr) ^2 + \biggr(\tilde{y}_i + \sum_{p\in
c(i)}\alpha_{pi} v_p - \tilde{y}_j - \sum_{p\in c(j)}\alpha_{pj}
v_p\biggr)^2 \biggr]~,
\end{multline}
where $\alpha_{pi}$ are the bilinear interpolation
coefficients.
\par
\begin{figure}[h]
\vbox{\center\includegraphics[width=6.5cm]{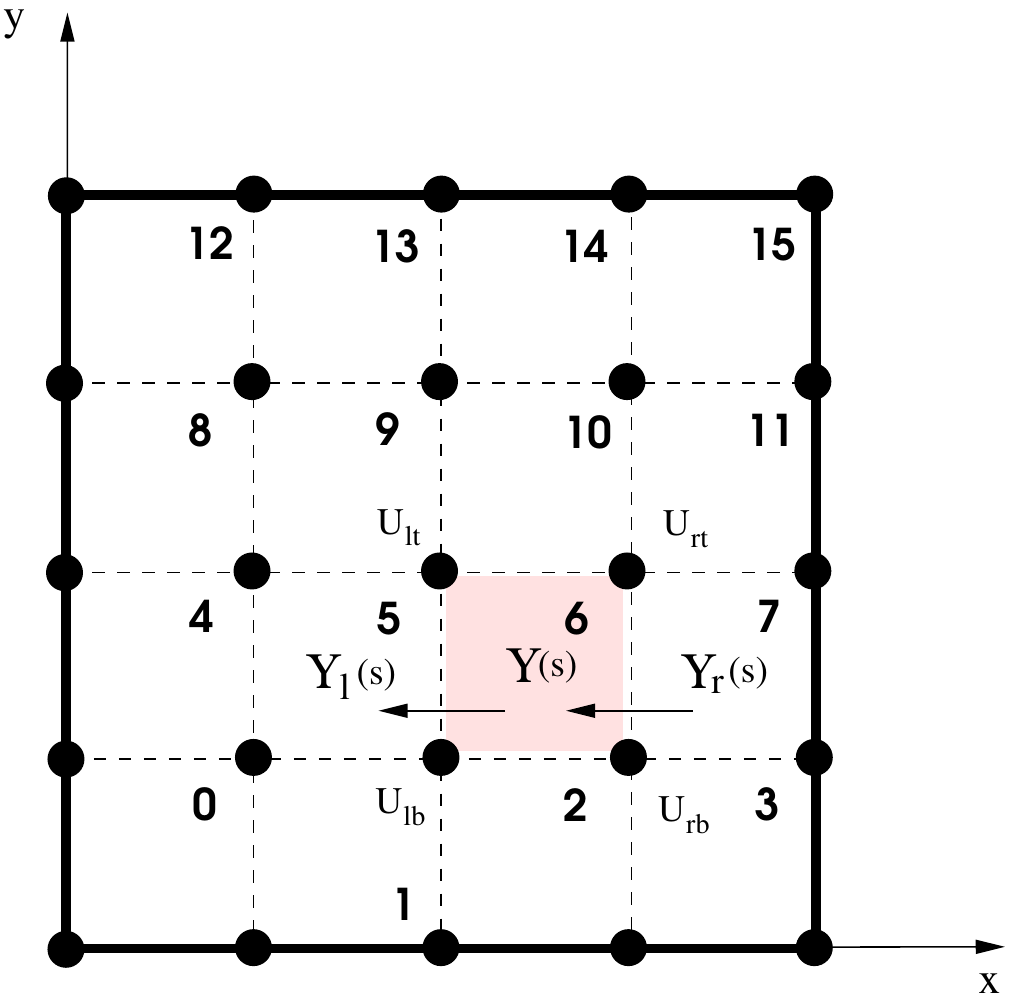}}
\caption{The horizontal direction flows of area considered for the
square $s$ (colored by gray) in the equidensity constraint
(\ref{ed-constraint}).}\label{circul-fig}
\end{figure}

 The reformulation of the equidensity constraint in
terms of the displacement variables relies on the rule of
conservation of areas. The initial total amount of vertex areas at
each square equals the current actual amount of areas dictated by
$(\tilde{x},\tilde{y})$. To estimate the amount of areas flowing
inside and outside a given square induced by the $u$ and $v$
displacements, we assume the nodes are evenly distributed inside
the squares. Under this assumption it is easier to estimate the
amount of area being transferred between two adjacent squares as
explained below. Consider, for example, a square $s$. Denote by
$\sar_{r(l,t,b)}(s)$ the total area of nodes overlapping with its
{\it right (left, top, bottom) neighbor} square. Let
$u_{rt(rb,lt,lb)}(s)$ be the $u$ values at the {\it right-top
(right-bottom, left-top, left-bottom)} corner of $s$ as shown in
Figure \ref{circul-fig}. To estimate the amount of areas entering
$s$ from the right we first calculate the average area (per
squared unit) in {\it both} squares: $(\sar(s)+\sar_r(s))/2\ar$.
We have to multiply this by the actual entering area (of nodes),
which is a rectangle of height $h_y$, the length of the border
between the two squares, and width, which is the average of the $u$
displacement at the middle of that border, namely,
$(u_{rt}+u_{rb})/2$. Thus the overall contribution of area from
the right is approximated by
\[
\frac{\sar(s)+\sar_r(s)}{2\ar} \cdot h_y \cdot
\frac{u_{rt}(s)+u_{rb}(s)}{2}~~~.
\]
A similar term is calculated at the left, and with $v$ instead of
$u$ also at the top and bottom. Note that if the assumed direction
of flow is wrong the resulting displacement will just turn out to
be negative.
\par The entire constraint for a square $s$ stating that the net flow of areas into the square
 should be equal to or be smaller than some demand $M(s)$ minus the current area in $u$, is given below:
\begin{multline}\label{ed-constraint}
\mathfrak{eqd}(s) = \frac{\sar(s) + \sar_r(s)}{2\ar}h_y
\frac{u_{\text{rt}}(s)+u_{\text{rb}}(s)}{2}- \frac{\sar(s) +
\sar_l(s)}{2\ar}h_y
\frac{u_{\text{lt}}(s)+u_{\text{lb}}(s)}{2} + \\
\frac{\sar(s) + \sar_t(s)}{2\ar}h_x \frac{v_{\text{rt}}(s)+
v_{\text{lt}}(s)}{2}- \frac{\sar(s) + \sar_b(s)}{2\ar}h_x
\frac{v_{\text{rb}}(s)+v_{\text{lb}}(s)}{2} \leq M(s) - \sar(s) ~.
\end{multline}
Next, to enforce the natural boundary conditions on $u$ and $v$,
namely, to forbid flows across the external boundaries, we simply
nullify all corresponding $u_p$ on the right and left boundary
points $\bugr$, and $v_p$ on the bottom and top boundary points
$\bvgr$.
Then the entire constrained optimization problem in terms of $u$
and $v$ and the initial approximation $(\tilde{x},\tilde{y})$ is
given by
\begin{equation}\label{full2var-problem}
\begin{array}{ll}
\mbox{\textbf{minimize}}  & \mathfrak{E}(u, v) ~~~(\text{given by } (\ref{min-func})\\
\mbox{\textbf{subject to}} &  \mathfrak{eqd}(s)=(\leq) M(s) - \sar(s) ~,~  \forall s \in \sgr ~ ;\\
                    & \mbox{if } p\in \bugr \text{ then } u_p=0 ~;\\
                    & \mbox{if } p\in \bvgr \text{ then } v_p=0 ~.
\end{array}
\end{equation}
We will simplify the formulation of (\ref{full2var-problem}) by
the concatenation of the two vectors $u$ and $v$ into one
$\bu=[\{u_i\}_{i=0}^{|\pgr|-1}~|~\{v_i\}_{i=0}^{|\pgr|-1}]$. We
will also omit the boundary conditions by directly replacing all
variables in $\bugr \cup \bvgr$ by $0$, and so, from
now on, we will refer to $\mathfrak{E}$ as
\begin{equation}\label{E-with-q-and-l}
\mathfrak{E}(\bu) = \frac{1}{2}\sum_{i,j}q_{ij}\bu_i \bu_j +
\sum_{i }l_i\bu_i + C~~~,
\end{equation}
where $i,j$ run over all the indices in $\bu \setminus \bugr
\setminus \bvgr$, $C$ is a constant and $q_{ij}$,
$l_i$ are the coefficients calculated directly from the previous
definition (\ref{min-func}) of $\mathfrak{E}$. Similarly rewrite
each $\mathfrak{eqd}(s)$ in (\ref{ed-constraint}) as
\begin{equation}\label{eqd-with-a-and-b}
\mathfrak{eqd}(s) = \sum_{i}a_{si}\bu_i=(\leq) b_s,
\end{equation}
where $b_s=M(s) - \sar(s)$.
\par Denote by $\lambda_s$, $s\in \sgr$ the Lagrange multiplier corresponding to the equidensity constraint
of square $s$.
If all the constraints are equality ones, the
Lagrangian minimization functional is
\begin{equation}\label{lagmin}
\Lag(\bu,\lambda)=\mathfrak{E}(\bu) + \sum_{s\in  \sgr}
\lambda_s(\mathfrak{eqd}(s)-b_s)~~~.
\end{equation}
So, we are looking for a critical point of the Lagrangian function, which is expressed by the system of linear
equations
\begin{equation}\label{lag-matrix}
\nabla \Lag(\bu,\lambda)=
\begin{bmatrix}
\nabla_{\bu}\Lag(\bu,\lambda) \\
\nabla_{\lambda}\Lag(\bu,\lambda)
\end{bmatrix}
=0 ~~~ .
\end{equation}
There are at least two factors that may cause (\ref{lag-matrix})
to be singular. First, the rank of $\nabla \Lag(\bu, \lambda)$ is
always less than its size by at least 1. This arises from the
equations of equidensity constraints in (\ref{lag-matrix}): their
sum always equals zero. The reason is that under the
boundary constraints the total amount of in-flows is always equal
to the total amount of out-flows.
In fact, the second summand in (\ref{lagmin})
can be replaced by
\[
\sum_{s\in  \sgr} (\lambda_s+Z)(\mathfrak{eqd}(s)-b_s)
\]
for any $Z$ without changing the minimization of $\Lag$ since
\[
Z\sum_{s\in  \sgr} (\mathfrak{eqd}(s)-b_s)=0~.
\]
Thus, important are not the values of $\lambda_s$ but only their
{\it differences}, and the singularity can be treated by an
additional constraint, say, $\sum_s d_s \lambda_s=0$, where $d_s=1~
\forall s \in \sgr$ (the introduction of $d_s$ is necessary for
the recursion of the multilevel solver; see Section
\ref{sCoarsening-scheme}). The additional term in
$\Lag(\bu,\lambda)$ is $\eta \sum_s d_s \lambda_s$, where $\eta$
is a ``pseudo-Lagrange'' multiplier. The following proposition
(with $k=1$) motivates the non-singularity of $\Lag$ with $\sum_s
d_s \lambda_s=0$.
\begin{proposition}
Given a symmetric $n\times n$ matrix $A$, for which $rank(A)=n-k$, let $x_i,~ i=1,...,k$ be an orthgonal basis of the null space of
$A$, that is, $Ax_i=0$. Then the following block matrix $B$ is
nonsingular
\[
B=\left(
\begin{array}{c|c}
 A      & X \\
\hline
 X^T  & 0
\end{array}\right),
\]
where $X=(x_1,...,x_k)$ is an $n\times k$ matrix of rank $k$.
\end{proposition}
\vspace{3mm}
\begin{proof}
Let $y$ be any vector in $\mathbb{R}^{n+k}$. Denote by $y'$ the
first $n$ components of $y$ and by $y''$ the last $k$ components,
that is, $y=\left( {y'}\over{y''} \right)$. We will prove that if
$By=0$, then $y=0$. The vector $By$ can be written in the following
block form:
\[
By=\left(
\begin{array}{c}
 Ay'+Xy''\\
\hline
 X^Ty'
\end{array}\right).
\]
Multiplying $Ay'+Xy''=0$ by $X^T$ from the left implies that
$y''=0$, and hence $Ay'=0$ and $y'=\sum_{i=1}^k \alpha_i x_i$.
Substituting the last relation into each of the last $k$ rows of
$B$ implies $x_j^Ty'=x_j^T \sum_{i=1}^k \alpha_i
x_i=\alpha_jx_j^Tx_j=0$ and thus $\alpha_j=0$ for $j=1,...,k$
yielding $y'=0$. Since $y'=0$ and $y''=0$ we may conclude that
$y=0$ as needed.
\end{proof}


\par The second kind of singularity in (\ref{lag-matrix}) may appear from possible empty squares. This
can be treated by adding a summand to (\ref{lagmin}) that
minimizes the total sum of all corrections $\beta \sum_i
\bu_{i}^2$, that is,  adds a $2\beta$-term to the diagonal of
$\nabla_{\bu}\Lag$, where $\beta$ is small enough to cause only
negligible change in a solution. This will prevent the inclusion
of zero-rows in $\nabla_{\bu}\Lag$, while possibly also bounding the
size of each correction in the solver below.
\par To summarize, the pseudo-Lagrangian functional $\Lag$ for our correction problem with equality
constraint is
\begin{equation}\label{pseudo-Lag}
\Lag(\bu, \lambda, \eta)= \frac{1}{2}\sum_{i,j}q_{ij}\bu_i \bu_j +
\sum_{i}l_i\bu_i  + \beta \sum_{i} \bu_{i}^2 + \sum_{s\in \sgr}
\lambda_s(\sum_{i}a_{si}\bu_i-b_s) +~ \eta \sum_{s\in \sgr} d_s
\lambda_s~,
\end{equation}
leading to the following system of equations
\begin{equation}\label{full-ineq-problem-in-q-l-a-b}
\begin{array}{ll}
\frac{1}{2}\sum_{j}q_{ij}\bu_j + l_i  + 2\beta \bu_{i} +
\sum_{s\in \sgr} \lambda_s a_{si} =0  ~,~ & \forall i
~\text{s.t.}~ \bu_{i} \in \bu
\setminus \bugr \setminus \bvgr \\
\sum_{i}(a_{si}\bu_i-b_s) +~ \eta d_s =0 ~,~ &
\forall s \in \sgr ~ \\
\sum_{s\in \sgr} d_s \lambda_s =0 ~.&
\end{array}
\end{equation}
\par Since in real world situations the total area is usually {\it bigger}
than the total area of all the vertices, the redefined
minimization problem
under {\it inequality} constraints will generally have the form
\begin{equation}\label{full-ineq-problem}
\begin{array}{ll}
\mbox{\textbf{minimize}}  & \mathfrak{E}(\bu) ~~~~~~~~~~~~~~~~~~~~~~~~~~~~(\text{given by }  (\ref{E-with-q-and-l}))\\
\mbox{\textbf{subject to}}   & \mathfrak{eqd}(s)\leq b_s ~,~
\forall s \in \sgr ~~~(\text{given by } (\ref{eqd-with-a-and-b}))~.
\end{array}
\end{equation}
\subsection{Multilevel solver for problem (\ref{full-ineq-problem})}\label{sMultSolver2}
\par To solve the constrained minimization problem (\ref{full-ineq-problem}), we use
multigrid techniques: standard geometric coarsening, linear
interpolation, Correction Scheme for the energy minimization and
the Full Approximation Scheme for the equidensity inequality
constraints; all are presented in Section
\ref{sCoarsening-scheme}. In addition, we have developed a fast
window minimization relaxation as explained in Section
\ref{sRelax}. The multilevel cycle is schematically summarized in
Section \ref{sMultilevel-cycle} in Algorithm {\bf
2D-layout-correction}.

\subsubsection{Coarsening scheme}\label{sCoarsening-scheme}
\begin{figure}
\vbox{\center\includegraphics[width=10cm]{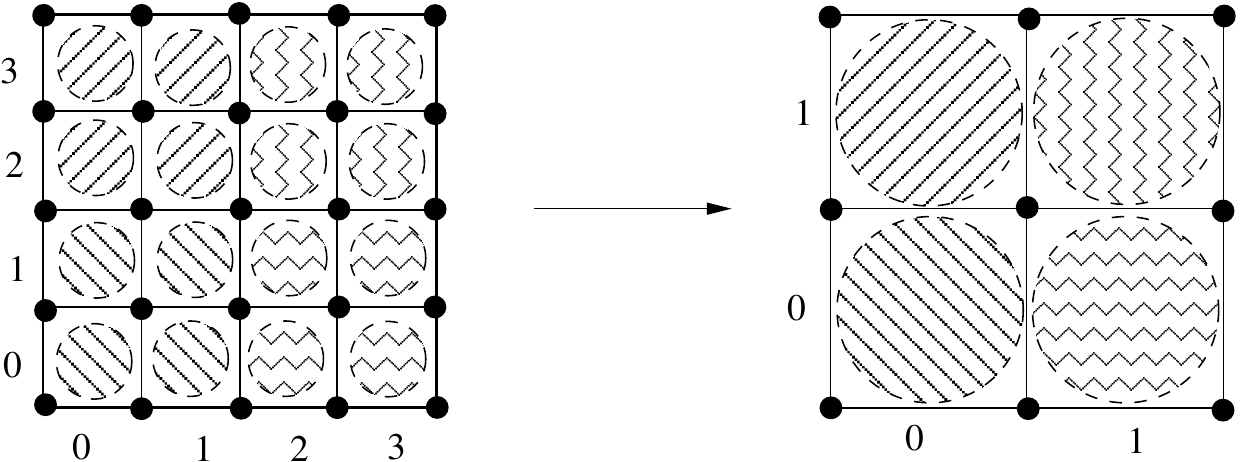}}
\caption{Geometric coarsening. The equidensity constraints of
every four similarly patterned squares at the fine level form one
equidensity constraint at the coarse level.}\label{geomgridcors}
\end{figure}

When the geometry of the problem is known we can choose a coarser
grid by the usual elimination of every other line, as shown in
Figure \ref{geomgridcors}. The correction computed at the coarse
grid points will be interpolated and added to the fine grid
current approximation. Let us introduce the notation
distinguishing between fine and coarse level variables and
functions. By lowercase and uppercase letters we will refer to the
variables, indexes, and coefficients of the fine ($\bu_i$, $i$,
$q_j$, etc.) and the coarse ($\bU_I$, $I$, $Q_J$, etc.) levels,
respectively. The subscripts $f$ and $c$ will be used to describe
the energy $\mathfrak{E}_{f}$ and $\mathfrak{E}_{c}$ and
pseudo-Lagrangian $\Lag_{f}$ and $\Lag_{c}$ functions at the fine
and the coarse levels, respectively.
\par Thus, the minimization part of the pseudo-Lagrangian (\ref{pseudo-Lag})
at the fine level is
\begin{align}\label{eqEf}
\mathfrak{E}_f=\frac{1}{2}\sum_{ij}q_{ij}\bu_i \bu_j+\sum_{i} l_i \bu_i~~~.
\end{align}
(Note that we have omitted the $\beta$ term from the following
derivation since it is merely an artificial added term.) Given a
current approximation $\tilde{\bu}$ of the fine level solution
$\bu$ and a correction function $\bU$ calculated at the coarse
level variables $\bU$, $\tilde{\bu}$ will be corrected by
\begin{equation}\label{equU}
\tilde{\bu}_i \leftarrow \tilde{\bu}_i+\sum_{I\ni i}
\alpha_{iI}\bU_I~,
\end{equation}
where the notation $\sum_{I\ni i}$ means that the sum is running
over all coarse gridpoints $p_I$ from which standard bilinear
interpolation is made to the fine gridpoint $p_i$.
\par Expressing the fine level energy
functional $E_f$ in terms of the coarse variables by substituting
(\ref{equU}) into (\ref{eqEf}) yields
\begin{align*}
\mathfrak{E}_f &=
\frac{1}{2}\sum_{ij}q_{ij}(\tilde{\bu}_i+\sum_{I\ni i}
\alpha_{iI}\bU_I)(\tilde{\bu}_j+\sum_{J\ni j} \alpha_{jJ}\bU_J)+\sum_i l_i(\tilde{\bu}_i+\sum_{I\ni i} \alpha_{iI}\bU_I)=\\
&=\frac{1}{2}\sum_{IJ} Q_{IJ}\bU_I\bU_J + \sum_{I} L_I\bU_I + C~,
\end{align*}
where ${Q}_{IJ}=\sum_{\substack{i\in I\\ j\in
J}}{q}_{ij}\alpha_{iI}\alpha_{jJ}$,
$L_I=\sum_{\substack{j\\i\in I}}{q}_{ij} \tilde{\bu}_j\alpha_{iI}+
\sum_{i\in I} l_i\alpha_{iI}~$ and $C$ is a constant. Thus, the coarse level energy
functional will be of the same structure as the fine level one,
namely,
\begin{equation*}
\mathfrak{E}_c = \frac{1}{2} \sum_{IJ} Q_{IJ}\bU_I\bU_J + \sum_{I}
L_I \bU_I~.
\end{equation*}
\par For each fine square $s$ the equidensity constraint $\mathfrak{eqd}(s)$ is
given by (\ref{eqd-with-a-and-b}).
The coarse equidensity constraints are constructed by merging
$2\times 2$ fine squares into one coarse square $S$.
The expression "$s\in S$" will refer to running over the four fine
squares $s$ that form the coarse square $S$ (see Figure
\ref{geomgridcors}). The $S$-th planar equidensity constraint  of
the coarse level (in the case of equality constraints only) is
obtained again by the substitution of (\ref{equU}):
\begin{eqnarray*}
\sum_{s\in S}\sum_i a_{si}\bu_i -\sum_{s\in S}b_s
& = & \sum_I A_{SI}\bU_I - B_S~,
\end{eqnarray*}
where $A_{SI}=\sum_{i\in I}\sum_{s\in S} a_{si} \alpha_{iI}$ and
$B_S=\sum_{s\in S}(b_s-\sum_i a_{si}\tilde{\bu}_i)$. Similarly (in
the case of equality constraints), the additional
$\eta$-constraint over all squares at the coarse level as
inherited from the fine level is $\sum_S D_S\Lambda_S=0$, where
$D_S=\sum_{s\in S}d_s$.
\par To complete the description of the coarse equations, we still need to transfer the equidensity {\it
inequality}
constraints.
For this purpose we will use the Full Approximation Scheme (FAS),
which is the general multigrid strategy applied to nonlinear
problems (see \cite{Brandt:1977:MLAa,vlsicad,mgbooktrott}). In
fact, there is no need for the FAS for the equality equidensity
constraints since it is a linear problem that can be solved by the
regular Correction Scheme (CS). The FAS-like coarsening rules are
needed and applied only on the set of equations derived from the
equidensity inequalities. Thus, our scheme is a combination of the
correction scheme for the energy equations derived from
(\ref{eqEf}) and (\ref{equU}) and FAS-like rules for the
equidensity equations.
\par To derive these equations we need to calculate the
{\it residuals} for both the fine and coarse grids. If $\Lag_f$ is
the pseudo-Lagrangian of the fine level system defined by
\begin{align}\label{eqLf}
\Lag_f=\mathfrak{E}_f+\sum_s \lambda_s(\sum_ia_{si}\bu_i-b_s)+\eta
\sum_s d_s\lambda_s~~~,
\end{align}
where $\mathfrak{E}_f$ is given by (\ref{eqEf}), then the
$\bu_i$-th residual of $\nabla \Lag_f$, where
$\tilde{\lambda}_s$ is the current value of the Lagrange
multiplier ${\lambda}_s$, is
\[
r_i^{\genr}=-l_i - \frac{1}{2}\sum_j q_{ij}\tilde{\bu}_j - \sum_s
\tilde{\lambda}_s a_{si}.
\]
Thus, the residual corresponding to the variable $\bU_I$ of
$\nabla \Lag_c$ (where $\nabla \Lag_c$ is the coarse level system
of equations analogous to (\ref{eqLf})) is
\begin{equation}\label{RI}
R_I^{\genr}=
\sum_{i\in I} \alpha_{iI}r_i^{\genr}~,
\end{equation}
where
$\alpha_{iI}$ are as in (\ref{equU}); that is, the fine-to-coarse
transfer is the adjoint of our coarse-to-fine interpolation. The
residual of the $s$-th equidensity constraint is
\[
r_s^{\geqd}=b_s - \sum_ia_{si}\tilde{\bu}_i - \tilde{\eta} d_s~,
\]
where $s$ runs over all fine squares and $\tilde{\eta}$ is the
current value of ${\eta}$. Therefore, the coarse equidensity
residual of square $S$ is
\begin{equation}\label{RS}
R_S^{\geqd}=
\sum_{s\in S}r_s^{\geqd}~.
\end{equation}
Finally the residual of the $\eta$-constraint is
\[
r_{\eta}=-\sum_s d_s\tilde{\lambda}_s=R_H.
\]

Denote by $LP(I)$ the linear part of the $\bU_I$-th equation in
the system $\nabla \Lag_c$
\[
LP(I)=\frac{1}{2}\sum_J Q_{IJ}\bU_J + \sum_S \Lambda_S A_{SI}.
\]
From the FAS rule for the $I$-th coarse equation stating that
$LP(I)=R_I^{\genr}+$ the current approximation of $LP(I)~$, we can
derive the $I$-th $\nabla \Lag_c$ equation
\begin{equation}\label{eqCoarseQL}
\frac{1}{2}\sum_J Q_{IJ}\bU_J + \sum_S \Lambda_S
A_{SI}-R_I^{\genr}-\frac{1}{2}\sum_J Q_{IJ}\bU_J^0 - \sum_S
\Lambda_S^0 A_{SI}=0,
\end{equation}
where $R_I^{\genr}$ is given by (\ref{RI}), $\bU_J^0=0$ and
$\Lambda_S^0=\frac{1}{4}\sum_{s\in S}\tilde{\lambda}_s$.
Similarly, the $S$-th square coarse equation for the equality
(inequality) constraint is
\begin{equation}\label{eqCoarseEqd}
\sum_IA_{SI}\bU_I+HD_S - R_S^{\geqd} - \sum_IA_{SI}\bU_I^0-H^0D_S
= (\leq)0~,
\end{equation}
where $R_S^{\geqd}$ is given by (\ref{RS}).
The last equation for the $H$-constraint is
\begin{equation}\label{eqCoarseEta}
\sum_S D_S\Lambda_S-R_H-\sum_S D_S\Lambda_S^0=0.
\end{equation}
Note that equations (\ref{eqCoarseQL}) to (\ref{eqCoarseEta}) are
the coarse grid equations analog to the system
(\ref{full-ineq-problem-in-q-l-a-b}). (A $2\beta \bU_I$ term may
be added to (\ref{eqCoarseQL}) for stability if needed.) The
correction received from the coarse level for the $\bu$ variables
is given by (\ref{equU}) and for the Lagrange multipliers
$\lambda$ by
\begin{equation}\label{eqLambdaInterpolation}
\tilde{\lambda}_s \leftarrow
\tilde{\lambda}_s+\Lambda_{\substack{S\ni
s}}-\Lambda_{\substack{S\ni s}}^0~.
\end{equation}
\subsubsection{Relaxation}\label{sRelax}
\par In our multigrid solver, as usual, the relaxation process is employed
as the smoother of the error of the approximation, before the
construction of the coarse level system and immediately
after interpolation from the coarse level. For this purpose we
have developed the {\it Window relaxation} procedure,
which extracts from the entire system small subproblems of
$m\times m$ squares and solves each separately, as explained
below. The running time of the entire relaxation process strongly
depends on the algorithm for solving one window. There exist many
 versions of well-known algorithms for the quadratic
minimization problem under linear inequality constraints (for a
survey see \cite{avriel}). However, since each window need be
solved only to a first approximation (because of the iterative nature
of the overall algorithm), in order to keep the running time low,
we have implemented a simple algorithm for approximately solving
each single window, as presented in {\bf SingleWindowSolver}.
\par 
Let
$\wnd = \{s\in \sgr | \text{ all squares within an $m\times m$
super-square}\}$ be a window of squares.
To solve the quadratic minimization problem in $\wnd$, we
fix at their current position all $\bu$ {\it outside} $\wnd$, as
well as all those that are on the boundary of $\wnd$ and
represent movement {\it perpendicular} to the boundary.
The minimization is done under the set of equidensity constraints
for the squares $s\in \wnd$.
The solution process for each single window is a simplified
version of the {\it active set method} and is iterative. At
each iteration $t$, for a given $\tilde{\bu}$ we first extract the
set (denoted by $S_t$) of squares for which the respective
inequality equidensity constraints are violated or almost
violated:
\[
S_t=\{s\in \wnd~|~ \eqd(s)> b_s -\epsilon \}~,
\]
where $\epsilon$ is positive and sufficiently small but not too
small to make $S_t$ numerically unstable (we have used
$\epsilon=0.0001*$(the square's area)). Then the inequality
constraints of $S_t$ are set to equalities ignoring the other
inequality constraints.
Let $\bpw$ be the set of all displacement indexes inside $\wnd$
(including those on the boundary of $\wnd$ directing {\it
parallel} to it).
For every $\bu_i, i \in \bpw$ we associate a correction variable
$\delta_i$ and we reformulate the pseudo-Lagrangian for $\wnd$ as a
functional of the $\delta_i$ variables as follows:
\begin{multline}
\Lag_{\wnd}(\delta, \lambda)= \frac{1}{2} \sum_{i,j\in
\bpw}q_{ij}(\tilde{\bu}_i+\delta_i)(\tilde{\bu}_j+ \delta_j) +
\frac{1}{2}\sum_{\substack{i\in \bpw ,~  j\not\in
\bpw}}q_{ij}(\tilde{\bu}_i+\delta_i)\tilde{\bu}_j +\\
\sum_{i\in \bpw}l_i(\tilde{\bu}_i+\delta_i)  +
 \beta \sum_{i\in \bpw} (\tilde{\bu}_{i}+\delta_i)^2 +
 \sum_{s\in  S_t}\lambda_s(\sum_{i\in \bpw}a_{si}\tilde{\bu}_i-b_s)~,
\end{multline}
where $\tilde{\bu}_i$ is the current value of $\bu_i$ and the
$\beta$ term is added for stability with $\beta=1$. Solving
$\nabla \Lag_{\wnd}(\delta,\lambda)=0$ we obtain the corrections
$\delta_i$ for $\tilde{\bu}_i, i \in \bpw$, which confine the
respective active set variables to the boundary of the equality
constraints manifold. However, while accepting this correction we
may violate other inequality constraints that were already
satisfied at the previous iteration $t-1$. Let us call this set of
new unsatisfied constraints $\overline{S}_t$. One way to overcome
this problem is to accept only a {\it partial} correction $\theta
\delta_i$, $i\in \bpw$, where $\theta$ is the {\it smallest}
number that brings some constraint from $\overline{S}_t$ to
equality. Accepting the correction $\theta \delta_i$ does not
violate any constraint from $\overline{S}_t$. At this point we
accept this partial correction and continue to the next iteration
$t+1$, excluding from the redefined $S_t$ the set of satisfied (by
equality) constraints from $S_t$ with negative Lagrange
multipliers $\lambda_s$. \vspace{6mm}

\begin{tabbing}
  \quad \=\quad \=quad \=quad \=quad         \kill
  {\bf SingleWindowSolver}($\wnd$, $\tilde{\bu}$)\\
  \keyw{begin}\\
  \> $t=0$ \\
  \> \keyw{Repeat} until "optimal enough" (explained at the end of Section \ref{sExamples})\\
  \>\> \keyw{If} $t=0$\\
  \>\>\>$S_t=\{$the violated equidensity constraints$\}$\\
  \>\> \keyw{Else}\\
  \>\>\>$S_t=\{$the violated equidensity constraints$\} \setminus $\\
            ~~~~~~~~~~~~~~~~~~~~$\{$those from iteration $t-1$ which satisfy equality and have $\lambda_s<0\}$\\
  \>\> \keyw{Solve} $\nabla \Lag_{\wnd}(\delta,\lambda)=0$ and extract the smallest $\theta$\\
  \>\> \keyw{Accept} the correction $\tilde{\bu}\leftarrow\tilde{\bu} +\theta \delta$\\
  \>\> $t\leftarrow t+1$\\
  \keyw{end}\\
\end{tabbing}
\par To achieve corrections for all variables, we will cover by these windows the
entire area in red-black order \cite{mgbooktrott}. For
computational reasons we have chosen to apply this relaxation for
very small windows (of size $4\times 4$ squares). To minimize the
effects of the boundary constraints in the windows and to enforce
the equidensity constraints over different super-squares, we scan the
entire domain two more times: once with
half-window size shift in the horizontal direction and once in the
vertical. Thus the overall relaxation process covers the domain
three times.

\subsubsection{The multilevel cycle}\label{sMultilevel-cycle}
\par Having defined the window relaxation, the interpolation, and
the coarsening scheme, the multilevel cycle naturally follows.
Starting from the given approximation $(\tilde{x},\tilde{y})$,
discretize the domain by a standard grid on which the $\bu$
variables are initially defined. Construct the system of equations
(\ref{full-ineq-problem-in-q-l-a-b}), and solve for the $\bu$
variables as follows. After applying $\nu_1$ window relaxation
sweeps, define the coarser level equations for the coarser grid,
apply $\nu_1$ window relaxation sweeps there, and continue to a
still coarser level. This process is recursively repeated until a small
enough problem is obtained. Solve this coarsest problem directly,
and start the uncoarsening stage by interpolating the solution of
the coarse level to the finer levels followed by $\nu_2$ window
relaxation sweeps on the finer level. Repeat until the correction
to the original problem is obtained. This entire multilevel cycle,
usually referred to as the {\it V-cycle}, is summarized in
procedure {\bf V-cycle-correction} below, where the superscript
index refers to the level number. (We have used $\nu_1=\nu_2=3$).
\vspace{6mm}
\begin{tabbing}
  \quad \=\quad \=quad \=quad \=quad         \kill
  {\bf V-cycle-correction}($\gr^i$, $\bu^i$, $\mathcal{C}^i$, $\lambda^i$, $\nabla\Lag^i$)\\
  \keyw{begin}\\
  \> \keyw{If} $\gr^i$ is a small enough grid\\
  \>\> \keyw{Solve} the problem exactly\\
  \> \keyw{Else}\\
  \>\> \keyw{Set} $\bu^i=0$\\
  \>\> \keyw{Apply} $\nu_1$ {\it Window relaxation} sweeps\\
  \>\> \keyw{Construct} $\gr^{i+1}$ the coarse level grid\\
  \>\> \keyw{Define} $\mathcal{C}^{i+1}$ to be the set of equidensity constraints\\
  \>\> \keyw{Initialize} the system of equations $\nabla\Lag^{i+1}$ given by (\ref{eqCoarseQL})-(\ref{eqCoarseEta})\\
  \>\> \keyw{Initialize} $\bu^{i+1}$ and $\lambda^{i+1}$\\
  \>\> \keyw{V-cycle-correction}($\gr^{i+1}$, $\bu^{i+1}$, $\mathcal{C}^{i+1}$, $\lambda^{i+1}$, $\nabla\Lag^{i+1})$\\
\>\> \keyw{Interpolate} from level $i+1$ to level $i$ using (\ref{equU}) and (\ref{eqLambdaInterpolation})\\
    \>\> \keyw{Apply} $\nu_2$ {\it Windows relaxation} sweeps\\
   \keyw{Return} $\bu^i$\\
\end{tabbing}

\subsection{The Full MultiGrid external driving routine}\label{sExternalCycle}
\par The solution of (\ref{full-ineq-problem-in-q-l-a-b}) is primarily dependent on the chosen grid size.
To enforce equidensity at all scales, it can be used within the
Full MultiGrid (FMG) framework. This is done by using a {\it
sequence} of increasing grid sizes (progressively finer
meshsizes), while employing a small number of V-cycles for each
grid size. We emphasize that the original problem
(\ref{full-problem}) is highly nonlinear, while the system of
equations with corrections in term of the displacement $\bu$ is
linearized around the current solution $(\tilde{x},\tilde{y})$.
Therefore, only a small correction should actually be taken from
the $\bu$ displacements when these $(\tilde{x},\tilde{y})$ are
being updated. Then, a new linear system can be formulated around
the new solution to obtain a new correction, and so forth. Thus, by small
steps of corrections we solve the original nonlinear problem via
the corrections calculated from the linear system of equidensity
constraints. For instance, we have tried to employ grids of sizes:
2, 4, 8, ... up to a grid with number of squares comparable to the
number of nodes in the graph. For each grid size the corresponding
set of equations (in terms of the displacement $\bu$) is solved
either directly (for small enough grids) or by employing the
V-cycles described in Section \ref{sMultilevel-cycle}. In either
cases the obtained solution $\bu$ is interpolated back to the
$(x,y)$ variables, introducing the desired correction to the
original variables of the problem. Various driving routines can be
actually used: each chosen grid size may be solved more than once
(e.g., use grids 2, 2, 4, 4, 8,...); the entire sequence of grids
may be repeated (e.g., 2, 2, 4, 4, 8,... , 2, 2, 4, 4, 8,...),
and so on. (see Section \ref{sExamples} for examples). These parameters
should in fact be optimized for each application according to the
concrete needs of the model.
\par The entire algorithm for the two-dimensional layout correction
is summarized below in Algorithm {\bf 2D-layout-correction}, where
the superscript 0 refers to the current chosen grid size.
\vspace{3mm}
\begin{tabbing}
  \quad \=\quad \=quad \=quad \=quad         \kill
  {\bf 2D-layout-correction}(graph $G$, current layout $(\tilde{x},\tilde{y})$)\\
  \keyw{begin}\\
  \> \keyw{Apply} for a sequence of grid sizes\\
  \>\> \keyw{Construct and initialize } $\gr^0$, $\bu^0$\\
  \>\> \keyw{Define} $\mathcal{C}^0$ to be the set of equidensity constraints\\
  \>\> \keyw{Initialize} the system of equations $\nabla\Lag^0$\\
  \>\> \keyw{V-cycle-correction}($\gr^0$, $\bu^0$, $\mathcal{C}^0$, $\lambda^0$, $\nabla\Lag^0$)\\
  \>\> \keyw{Update } $(\tilde{x},\tilde{y})$ from $\bu^0$\\
   \keyw{Return} $(\tilde{x},\tilde{y})$\\
\end{tabbing}

\section{Examples of graph drawing layout correction}\label{sExamples}
\par
\begin{figure}
\begin{center}
$\begin{array}{cc}
\includegraphics[width=2in]{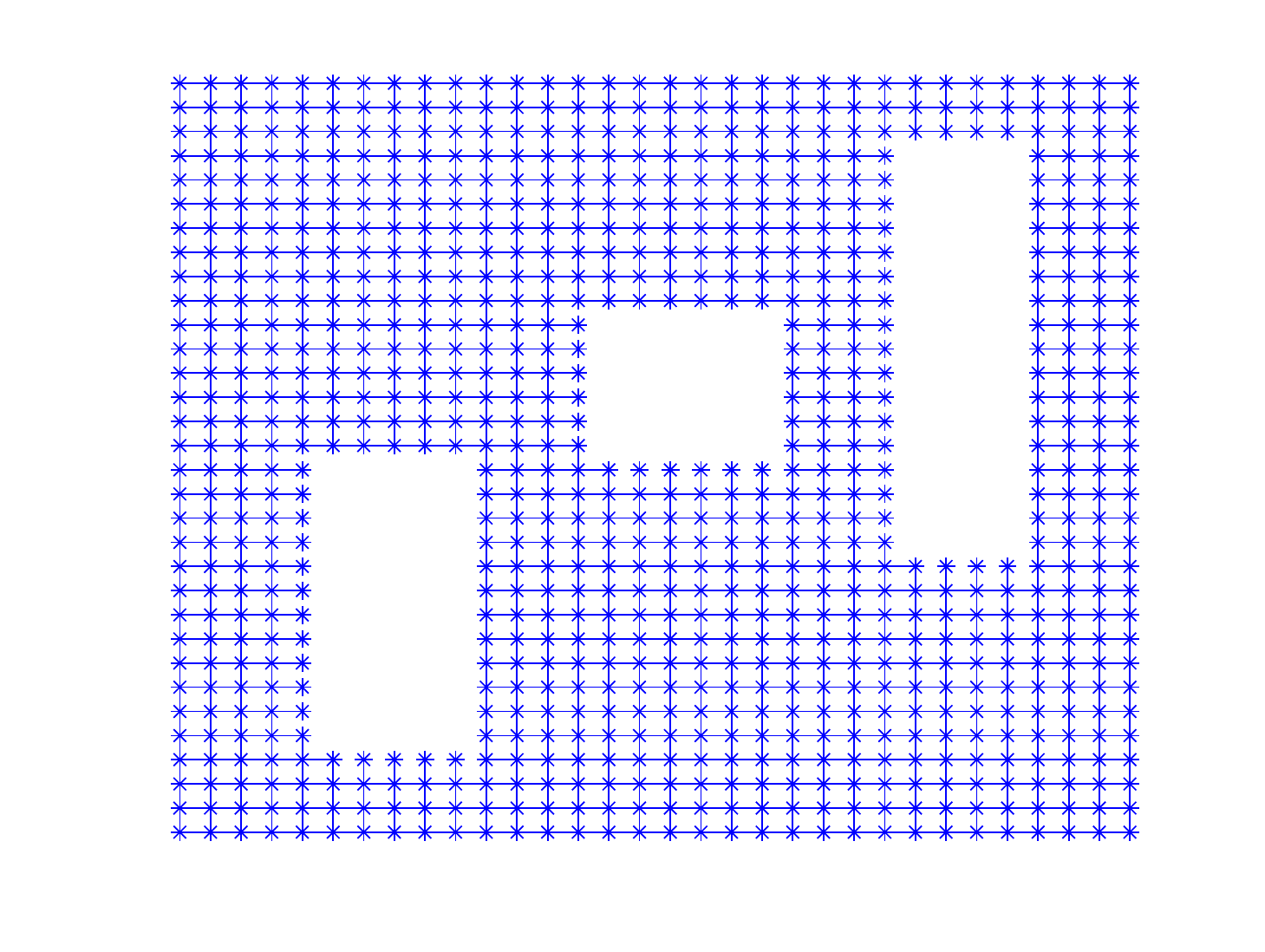} &
\includegraphics[width=2in]{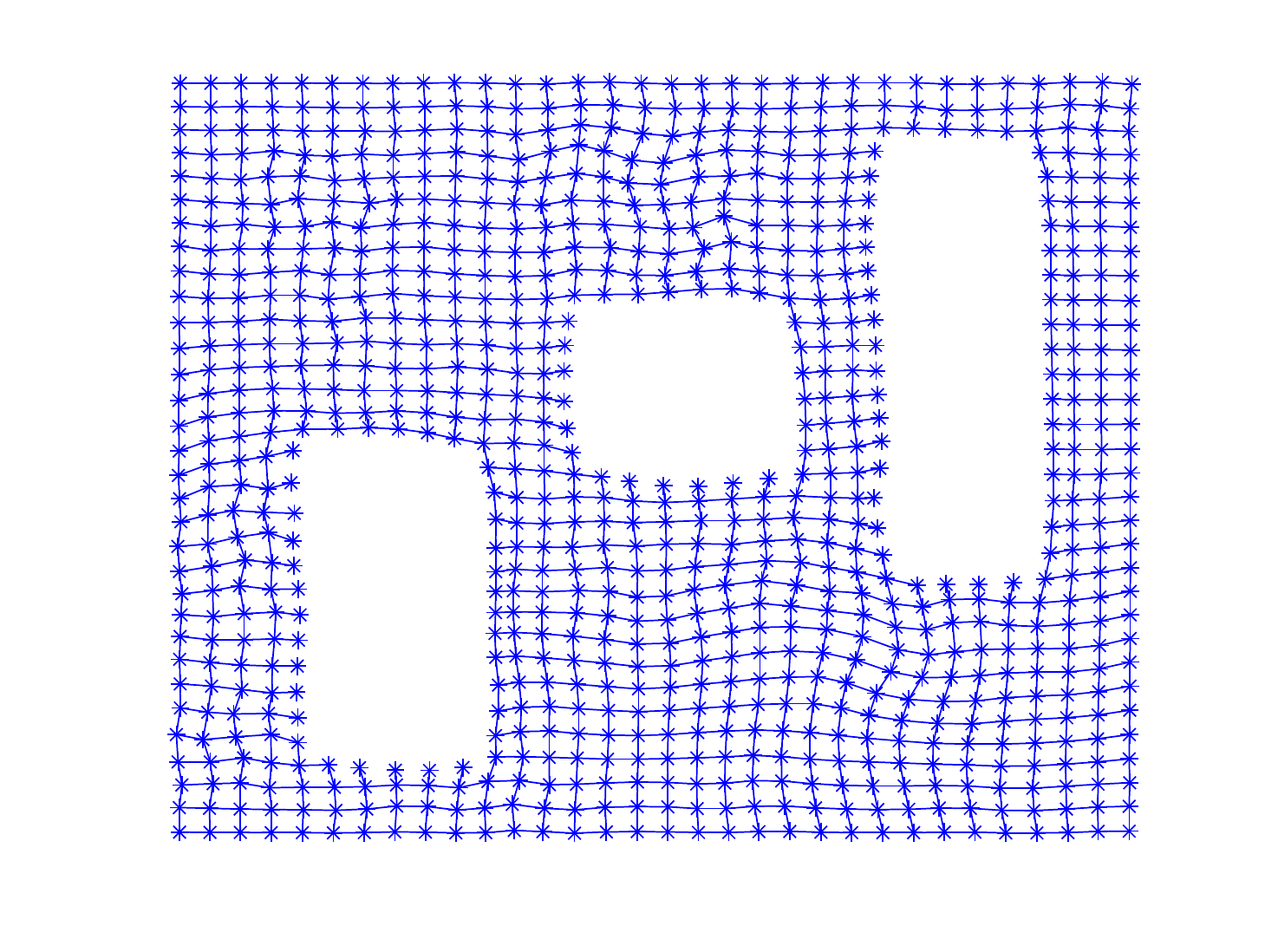} \\ [0.1cm]
\mbox{\bf (a1)} & \mbox{\bf (a2)}\\
\includegraphics[width=2in]{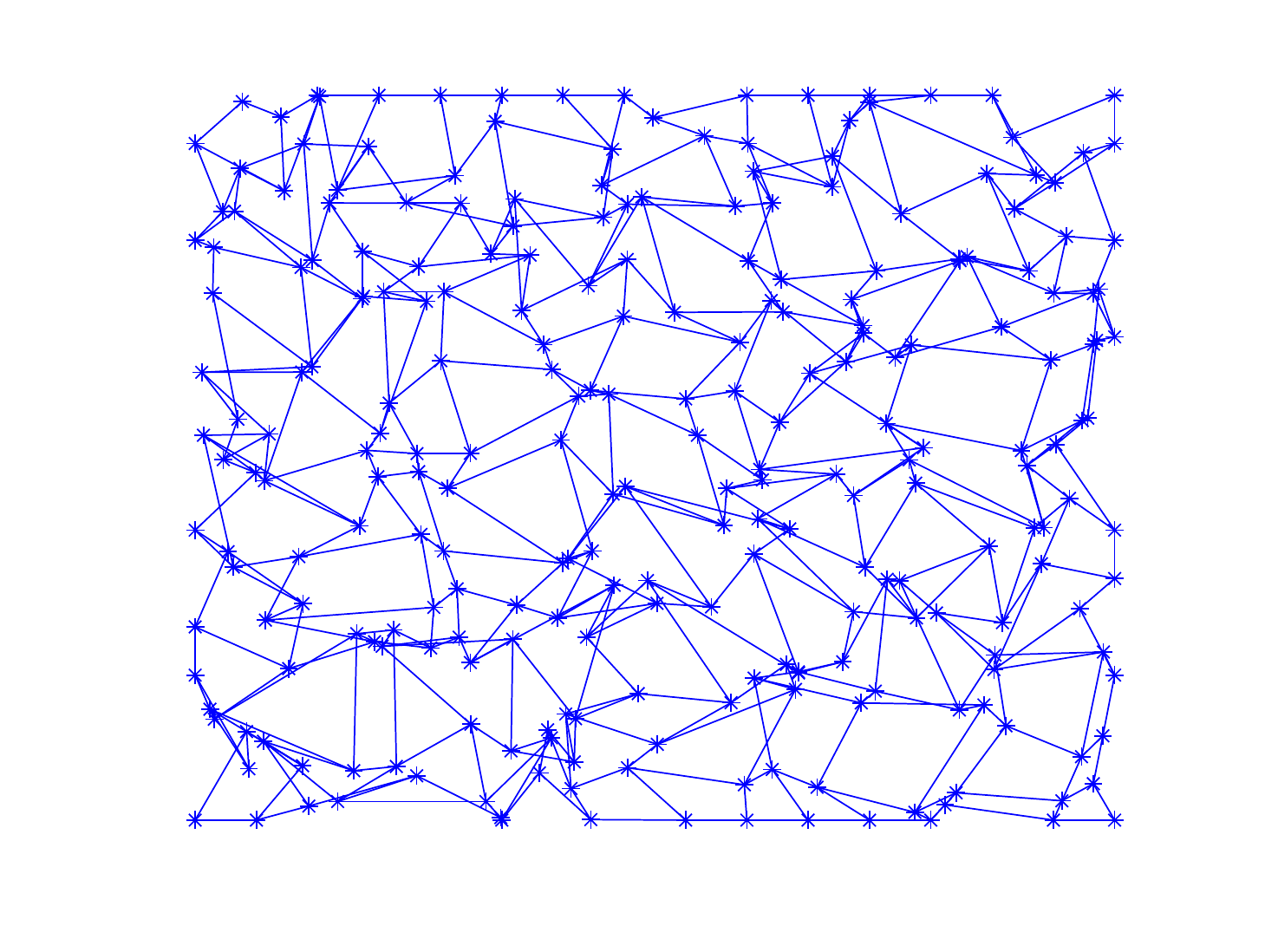} &
\includegraphics[width=2in]{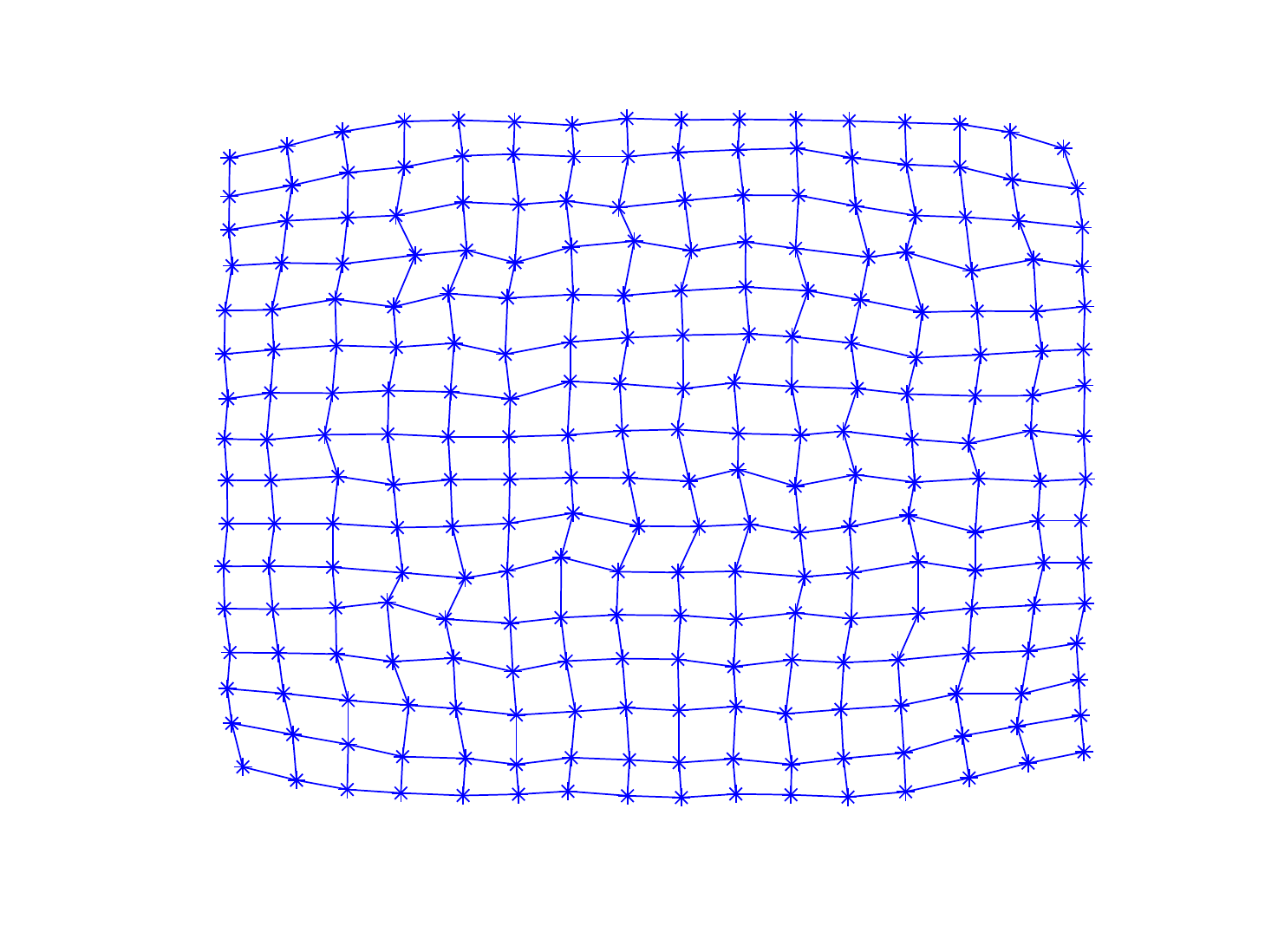} \\ [0.1cm]
\mbox{\bf (b1)} & \mbox{\bf (b2)}\\
\includegraphics[width=2in]{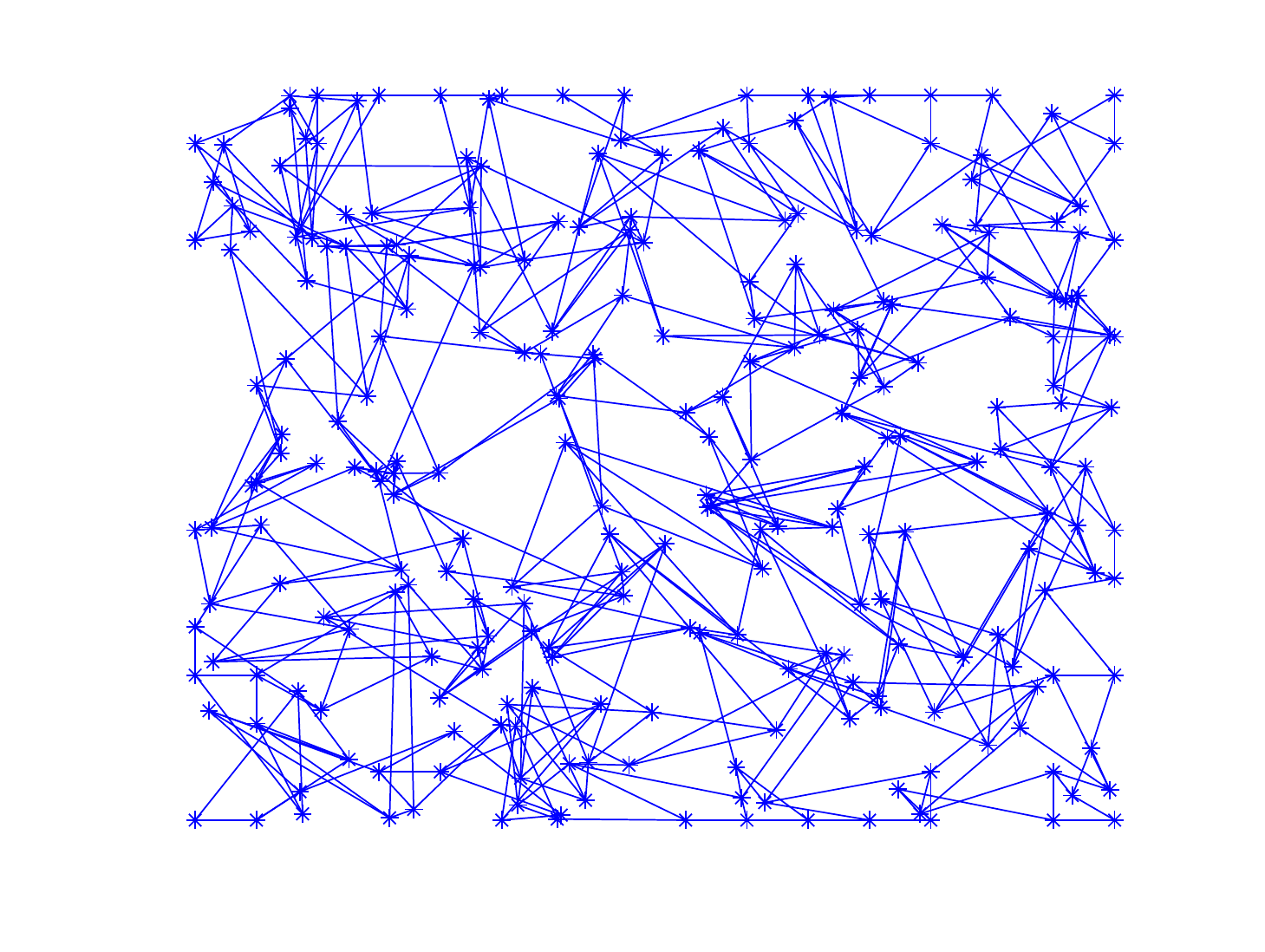} &
\includegraphics[width=2in]{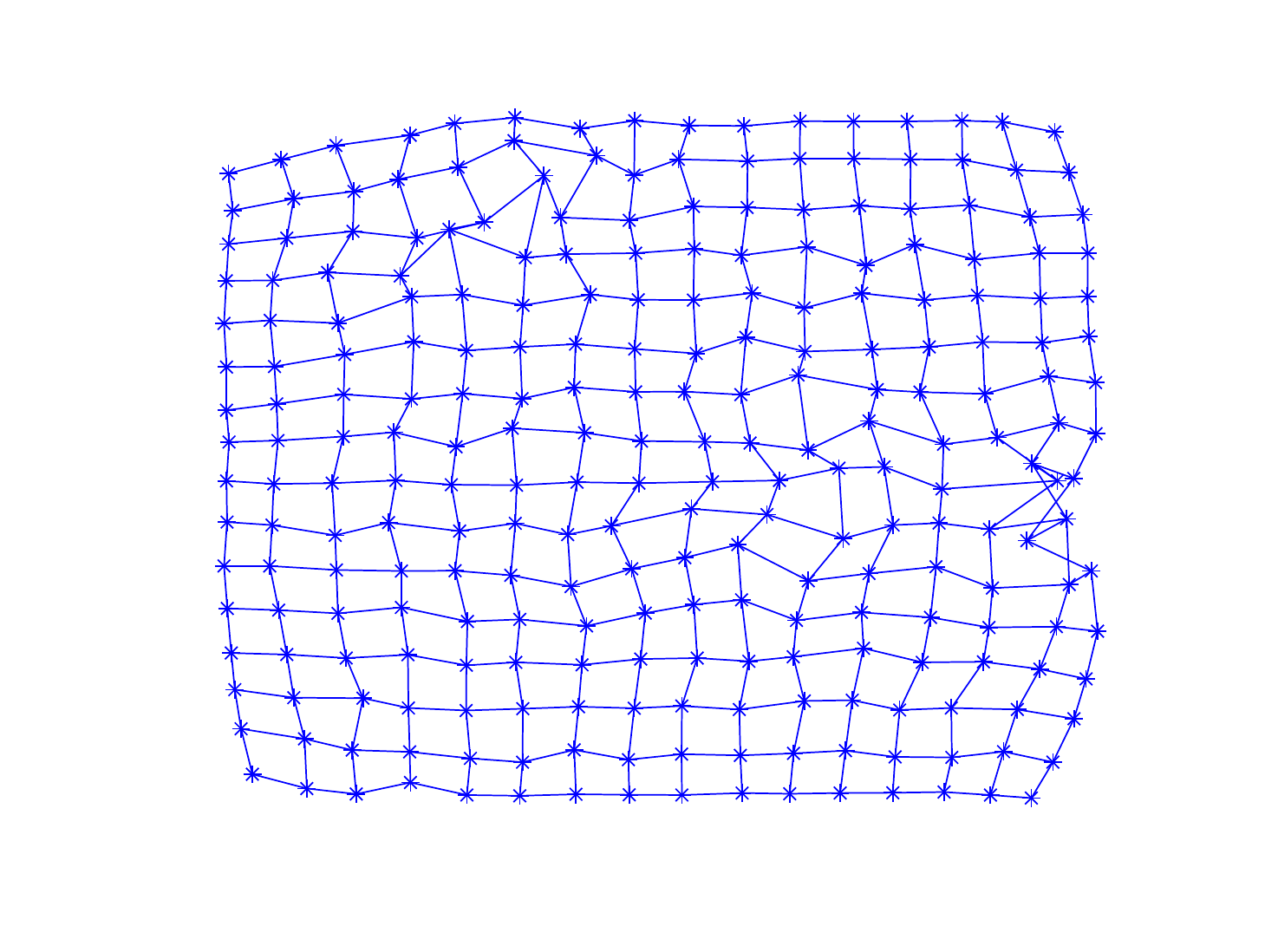} \\ [0.1cm]
\mbox{\bf (c1)} & \mbox{\bf (c2)}
\end{array}$
\end{center}
\caption{Examples of the 2D-layout of graphs with equal vertices.}
\label{figtest}
\end{figure}

\begin{figure}
\begin{center}
$\begin{array}{cc}
      \includegraphics[width=2.5in]{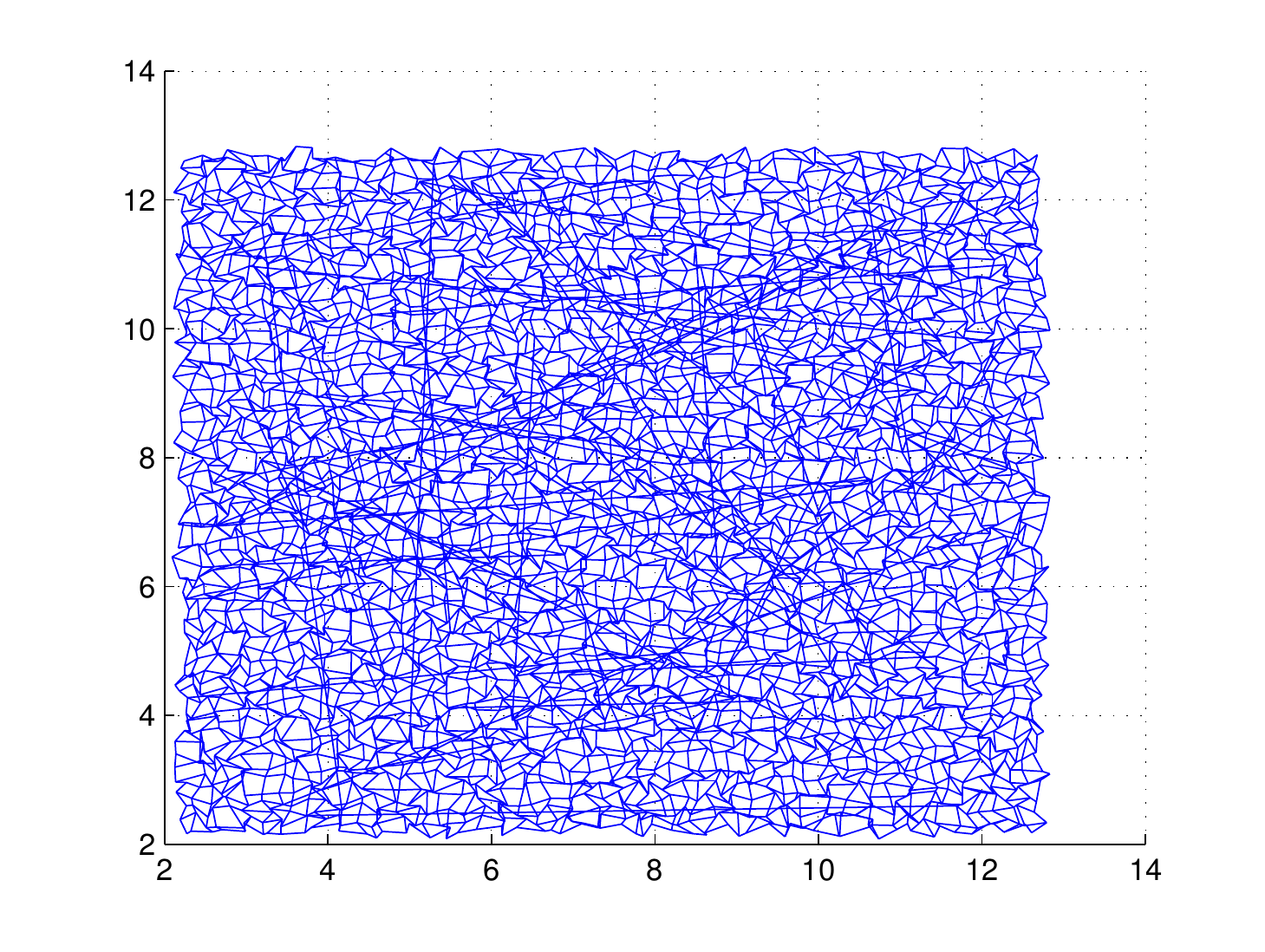} &    \includegraphics[width=2.5in]{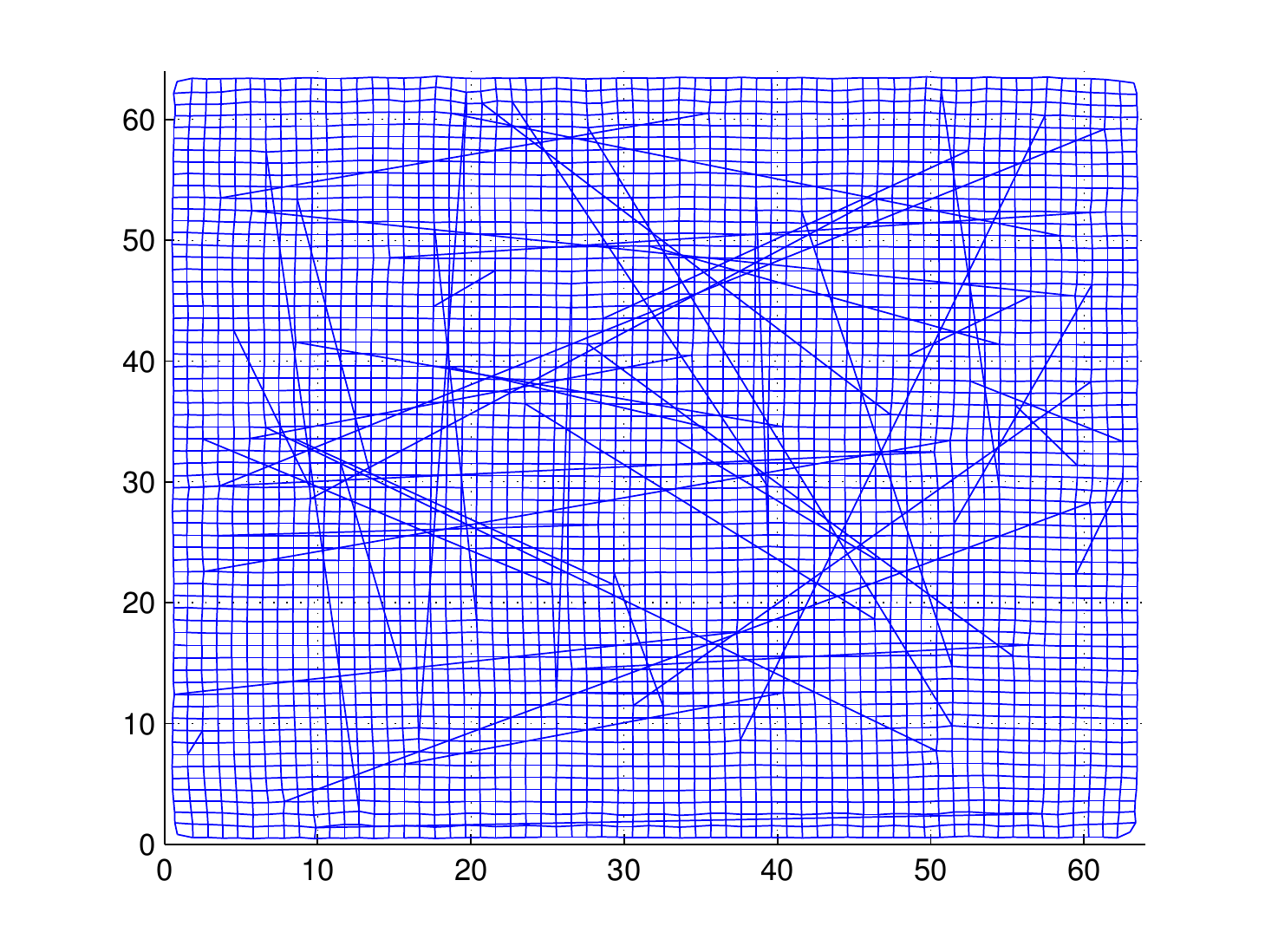}\\
      \textbf{(a)} &  \textbf{(b)}
 \end{array}$
 \end{center}
  \caption{Example of the layout of the $64\times 64$ mesh with additional random edges
  (note the different scales of the two figures):
  (a) starting from a compressed and perturbed configuration at the bottom-left corner,
  (b) the resulting picture using V-cycles.}
  \label{figm64add}
\end{figure}

\begin{figure}
\begin{center}
$\begin{array}{cc}
      \includegraphics[width=2.5in]{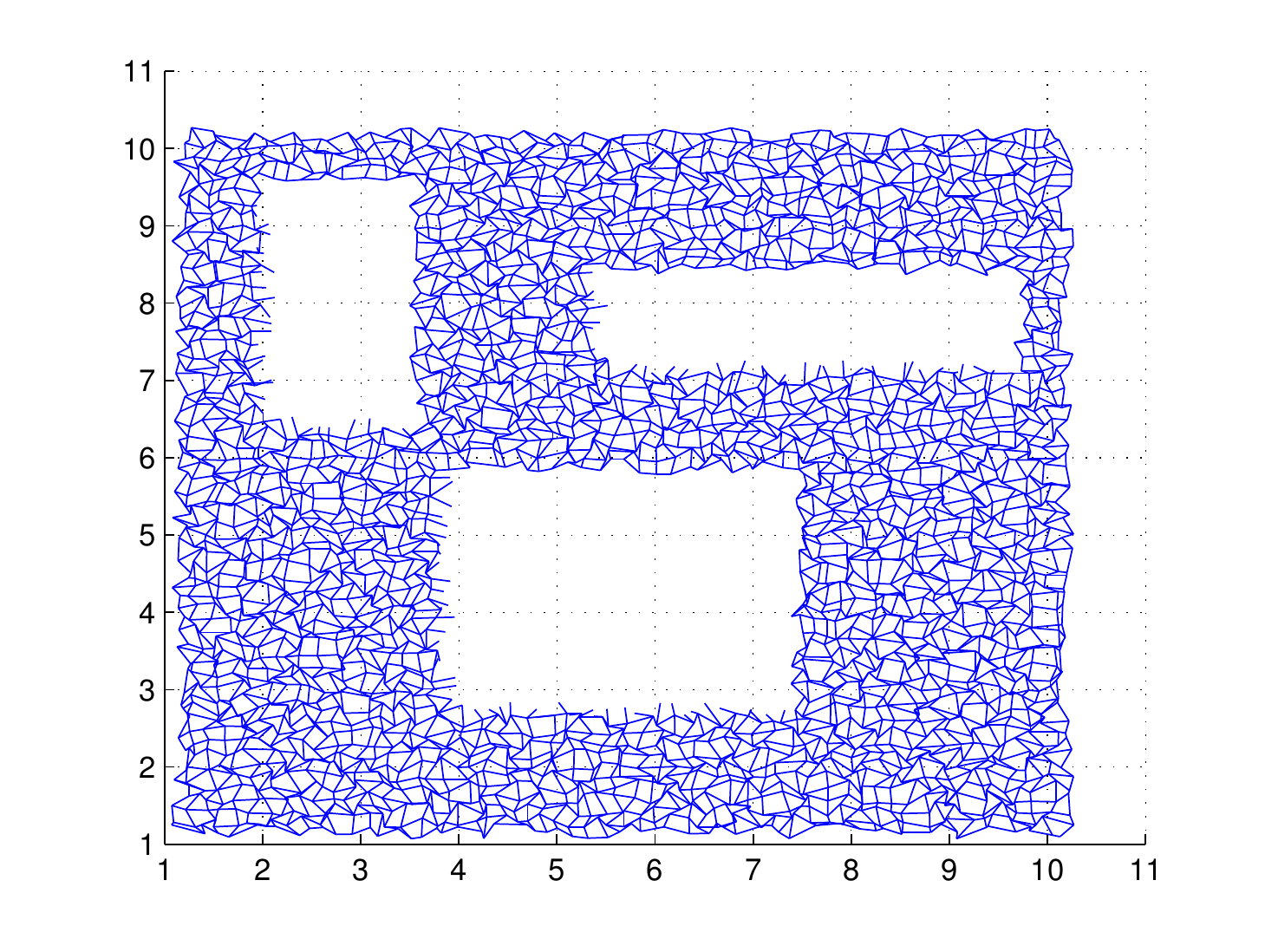} &    \includegraphics[width=2.5in]{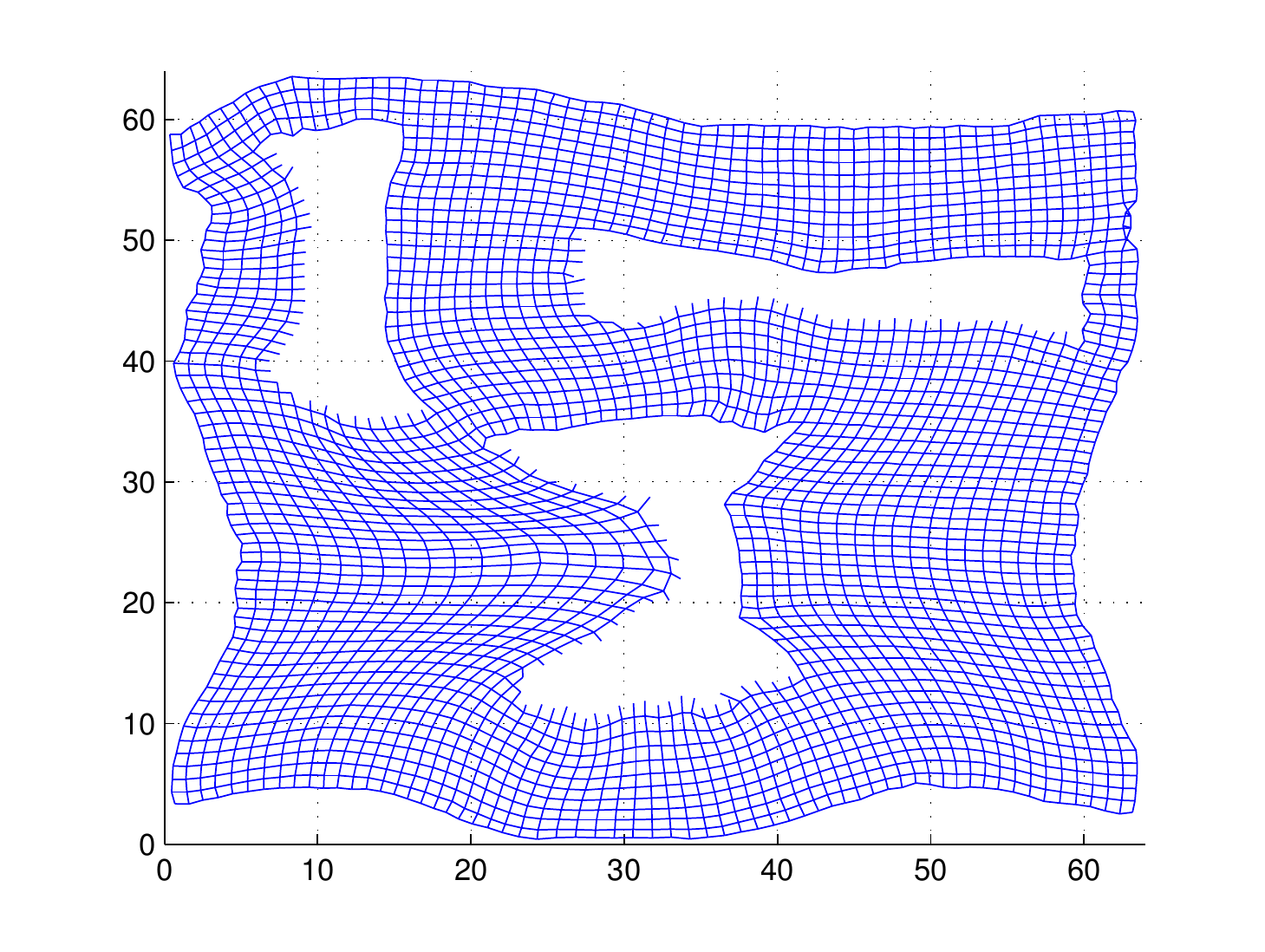}\\
      \textbf{(a)} &  \textbf{(b)}
 \end{array}$
 \end{center}
  \caption{Example of the $64\times 64$ mesh with three holes layout
  (note the different scales of the two figures):
  (a) starting from a compressed and perturbed configuration at the bottom-left corner, (b) the resulting picture using V-cycles.}
  \label{figm64hole}
\end{figure}

\begin{figure}
  \centerline{
    \mbox{\includegraphics[width=4.00in]{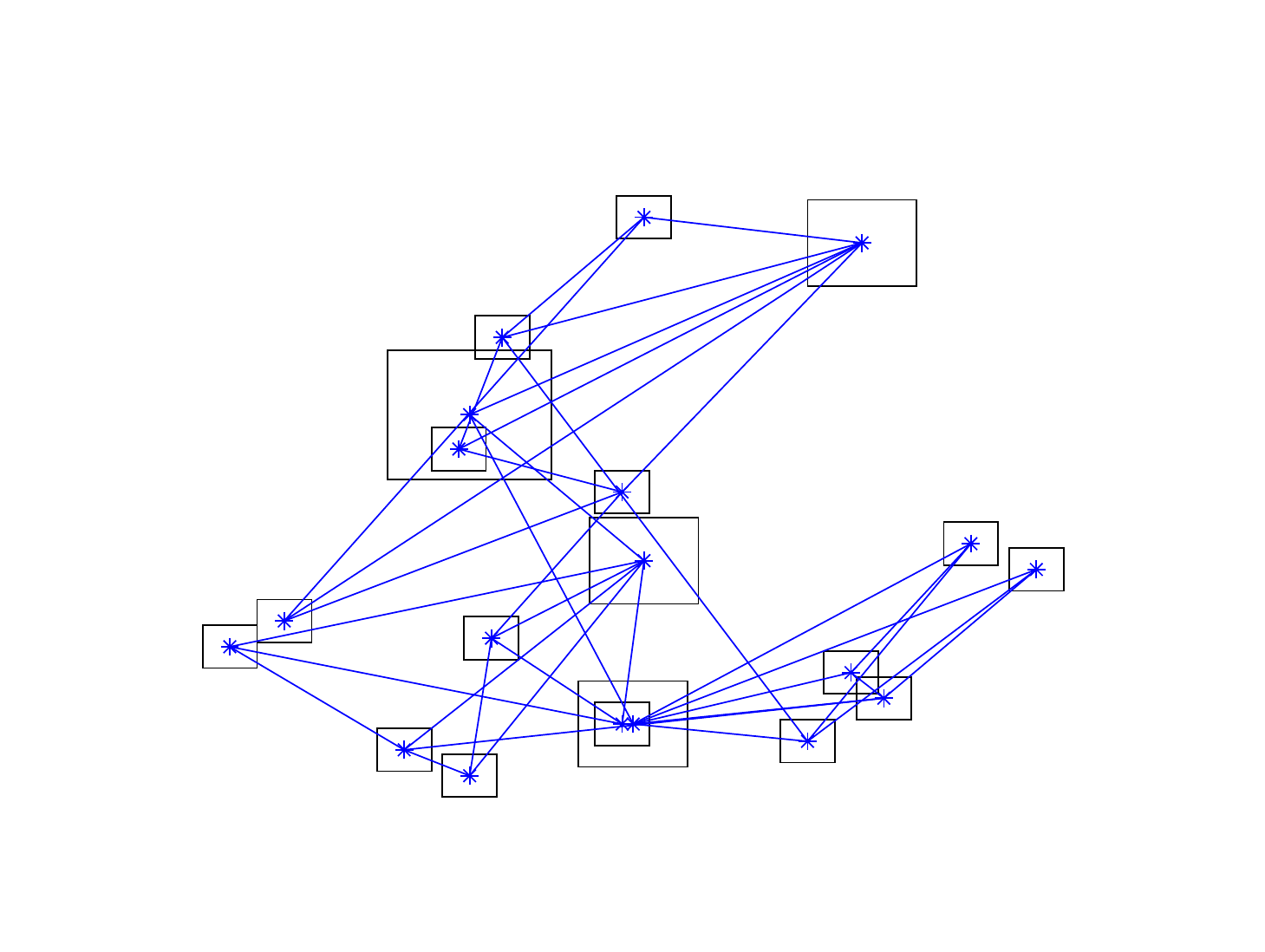}}
    \mbox{\includegraphics[width=3.00in]{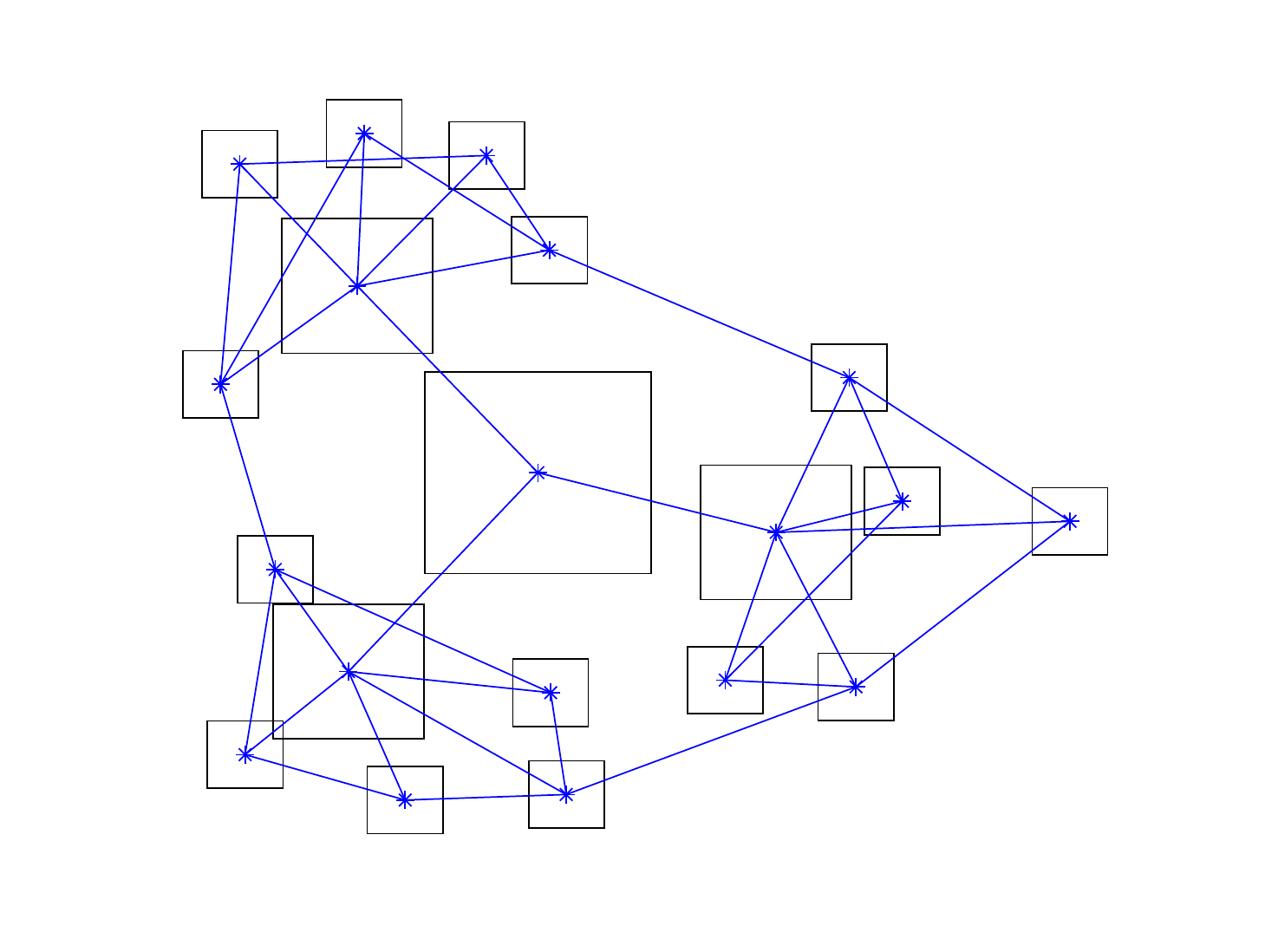}}
  }
  \caption{Example of the 2D-layout of a graph with nonequal volumes.}
  \label{figvoltest}
\end{figure}

\begin{figure}
  \centerline{
    \mbox{\includegraphics[width=3.00in]{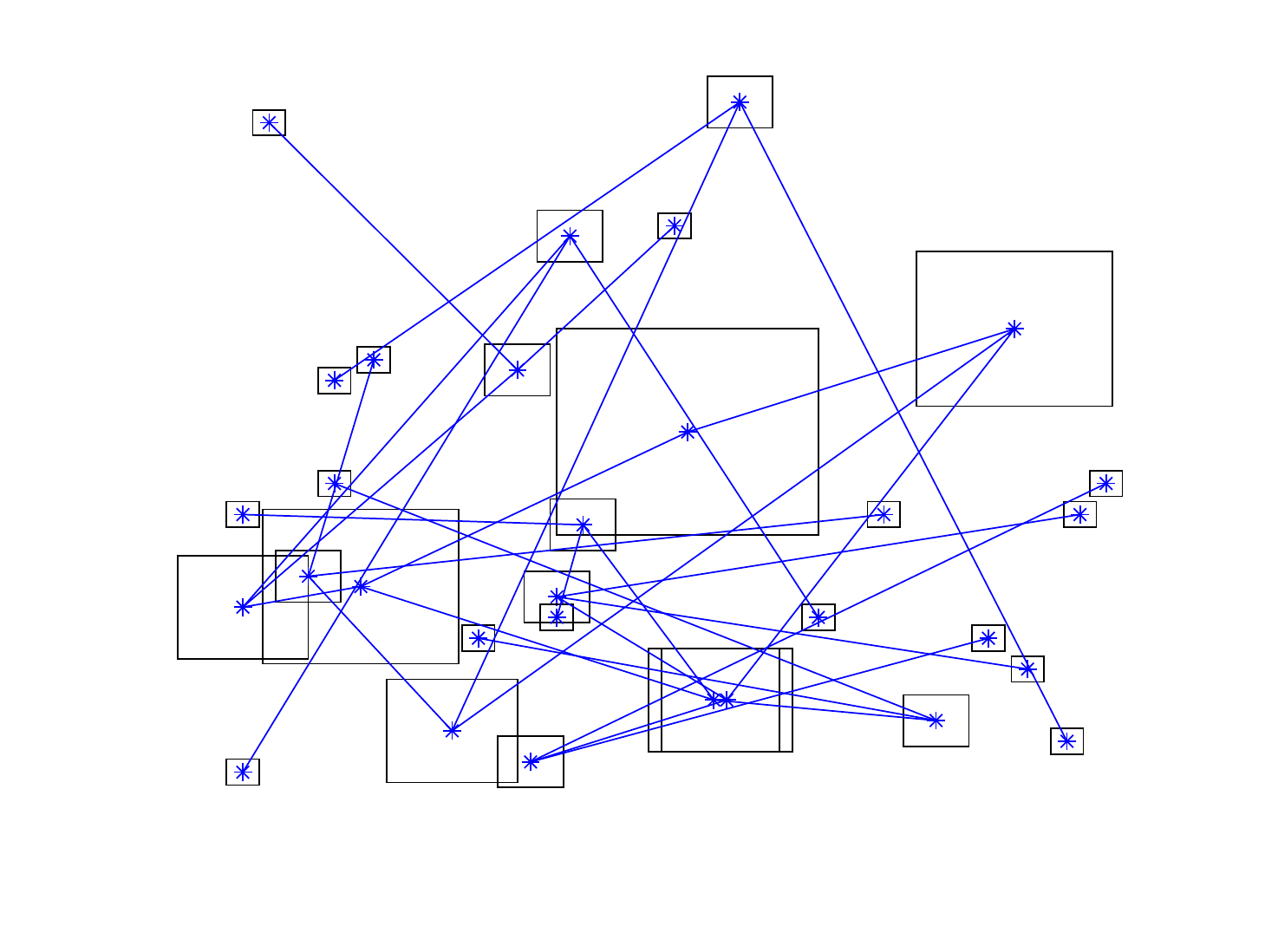}}
    \mbox{\includegraphics[width=3.00in]{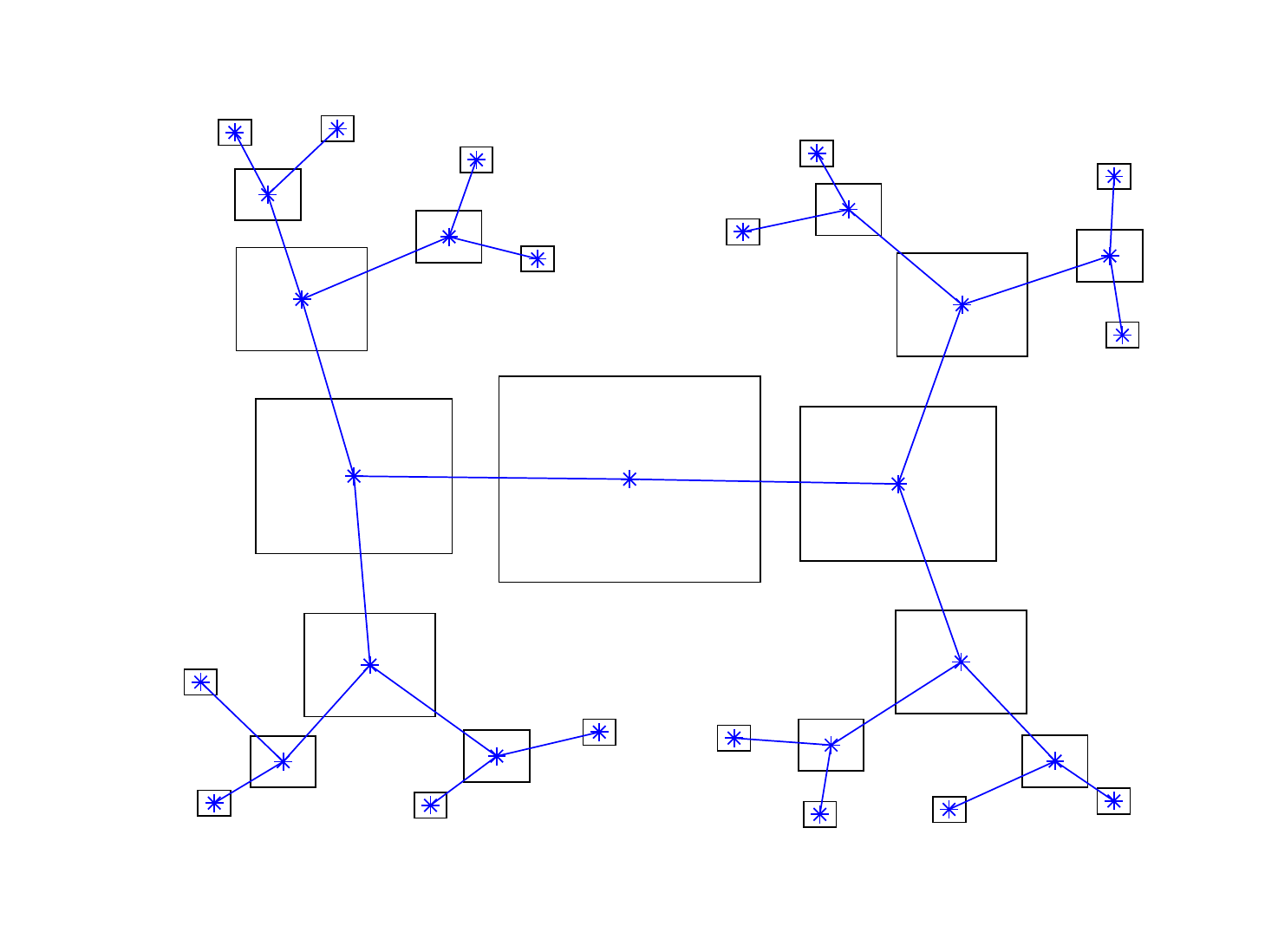}}
  }
  \caption{An example of the 2D-layout of a 5-level binary tree with non-equal vertices.}
  \label{figvoltest-tree}
\end{figure}

 As previously mentioned, the graph drawing problem is of interest
for many applications. Therefore, we have chosen to demonstrate
the abilities of our algorithm for this problem. In this section
we will present several results of the two-dimensional layout
correction algorithm using inequality constraints. The set of
examples is shown in Figures \ref{figtest} and \ref{figm64add} to
 \ref{figvoltest-tree}, each organized in two columns. The initial and
 final layouts of the graph are shown in the same row, in the left and
 the right columns, respectively.
Note that finalizing the "nice graph" representation of these
examples is beyond the scope of this work. The various
"beautifying" procedures used by different applications may, of
course, be used at the end of our cycles to enhance the
visualization results.

\par The first example consists
 of a mesh graph with three holes (Figure \ref{figtest}, row (a)).
 It is intended to demonstrate that the empty space stays empty and
 the energy is thus kept low. More complicated examples are shown
 in Figure \ref{figtest}, rows (b) and (c). The initial optimal positions
 of the mesh's vertices were randomly changed by independent shifts
 in different directions within a distance $d$:
\begin{equation*}
d~\leq~ \left\{
  \begin{array}{ll}
  2h_x & \text{ in example (b)}\\
  4h_x & \text{ in example (c)}~,\\
  \end{array} \right.
\end{equation*}
where $h_x$ is the length of a square on the initially taken 32x32
grid, such that the mesh size of the graph is actually $2h_x$. Let
us call these meshes $M_1$ and $M_2$, respectively. While the
correction of $M_1$ looks really nice, two switched vertices at
the right-hand side of $M_2$ demonstrate a weak point in our
algorithm that certainly must be improved by a local "beautifying"
procedure, which in general depends on the real application. The
initial layout (c1) is more complicated than (b1), while the
desired final layouts are similar.
\par
\begin{figure}
\vbox{\center\includegraphics[width=7cm]{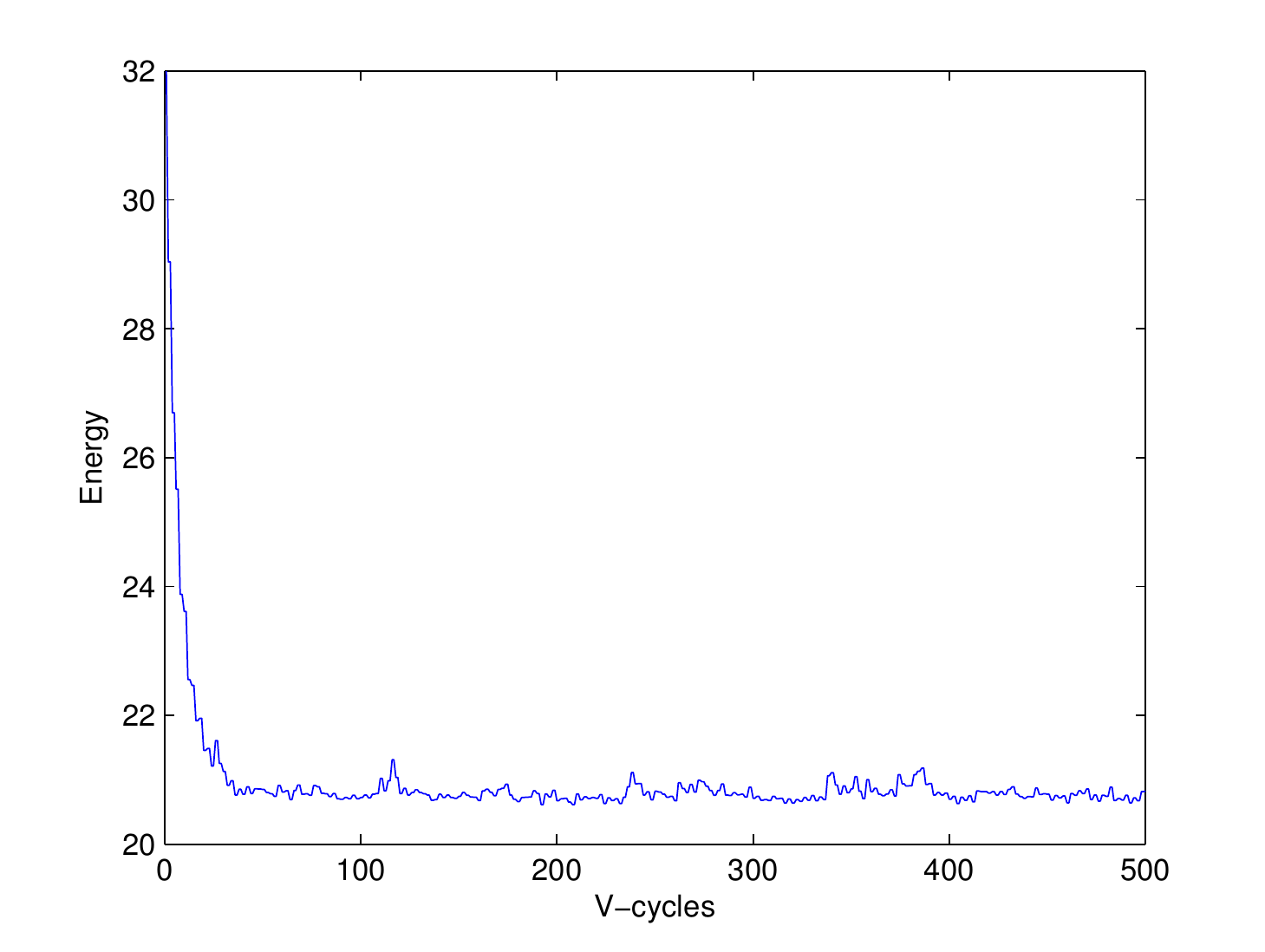}}
\caption{Energy behavior of the mesh at Figure \ref{figtest}-(c),
when employing complete V-cycles with $16\times 16$ and $32\times
32$ alternately.}\label{vc-conv}
\end{figure}

\begin{figure}
\vbox{\center\includegraphics[width=7cm]{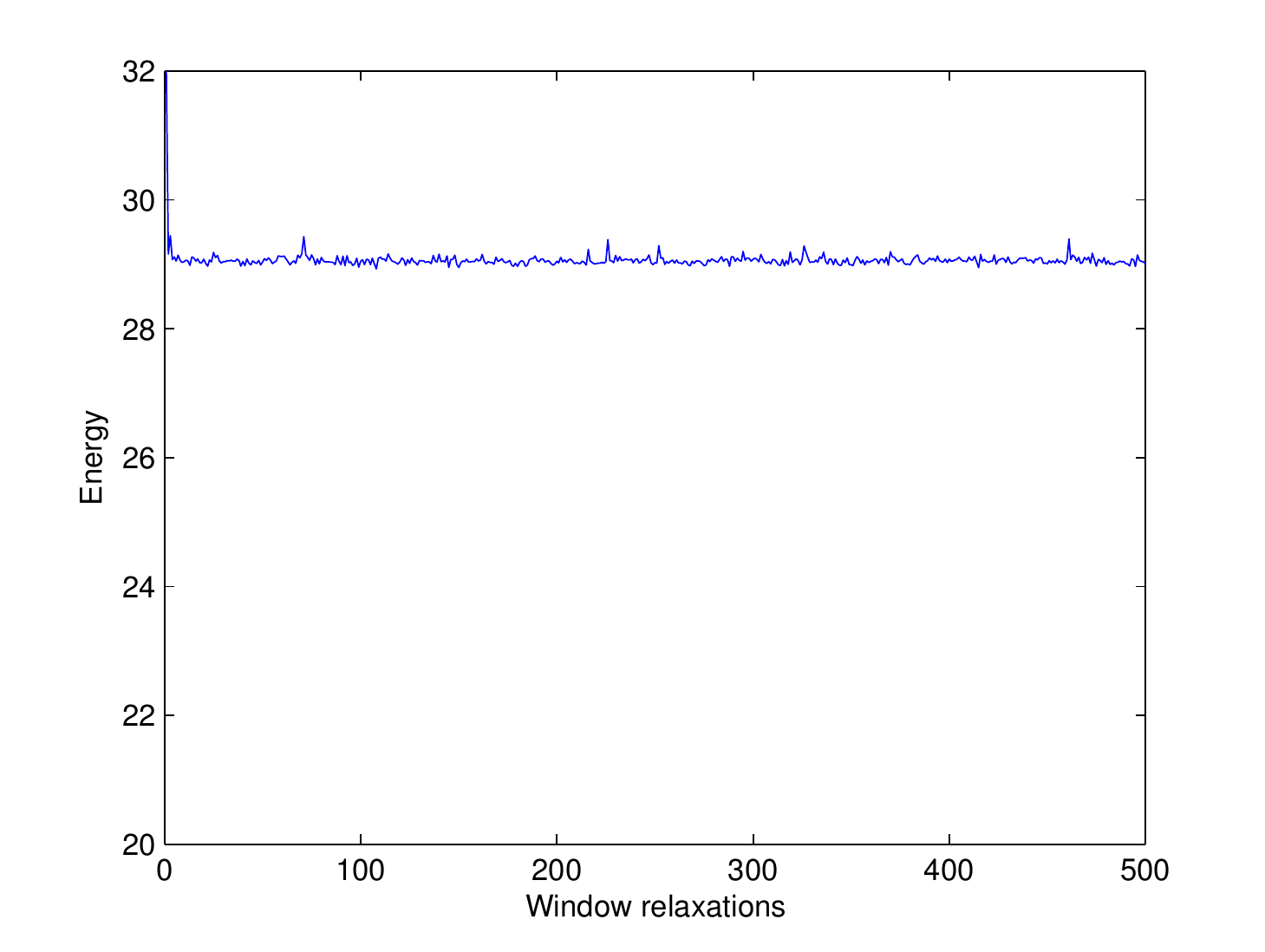}}
\caption{Energy behavior of Window relaxation iterations
($16\times 16$ grid) of the mesh at Figure
\ref{figtest}-(c).}\label{wr-conv}
\end{figure}

\begin{figure}
\vbox{\center\includegraphics[width=7cm]{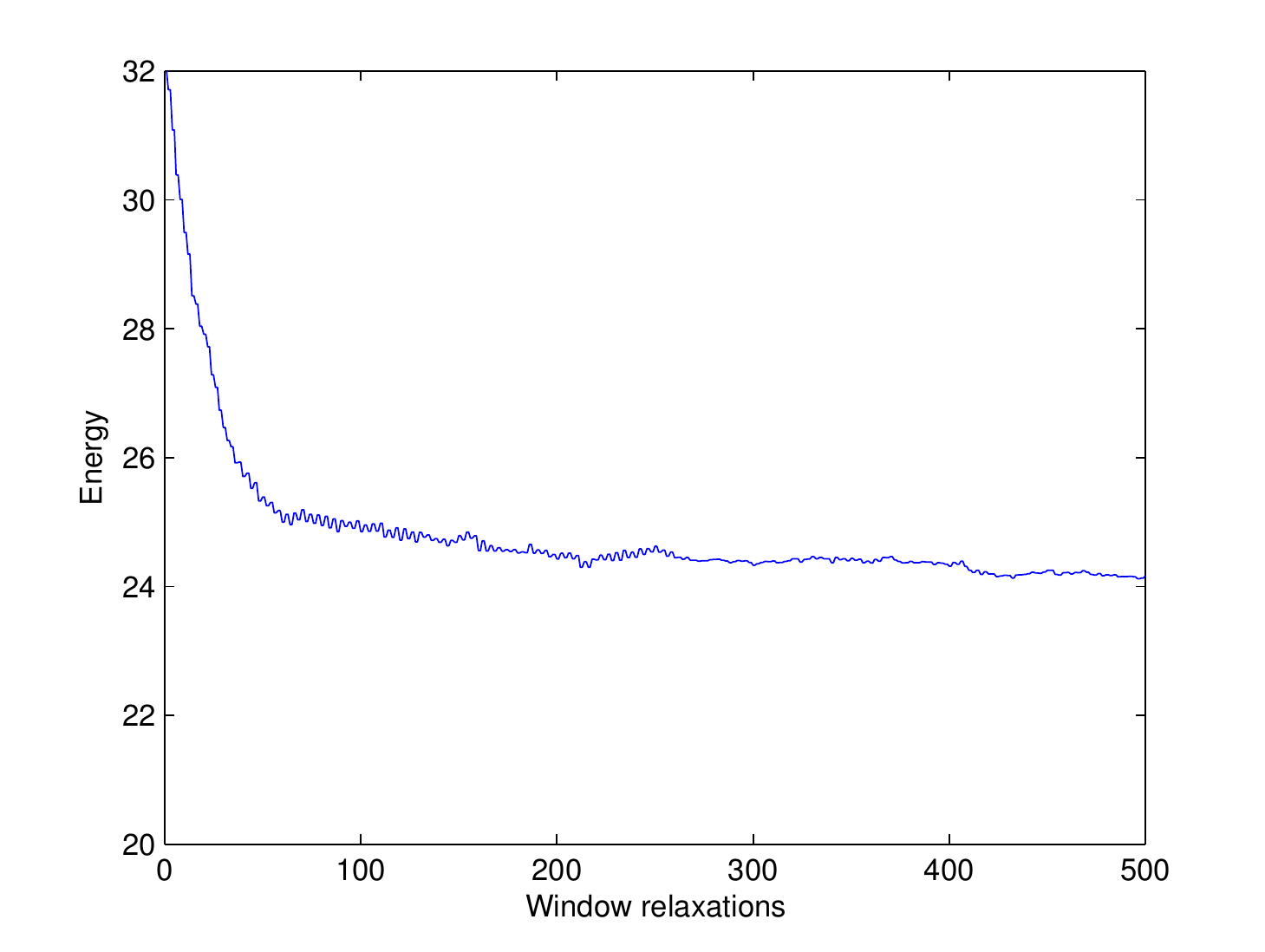}}
\caption{Energy behavior of Window relaxation iterations
($16\times 16$ and $32\times 32$ grids) of the mesh at Figure
\ref{figtest}-(c).}\label{wrb-conv}
\end{figure}

 A typical example of the energy behavior is presented
in Figures \ref{vc-conv}-\ref{wrb-conv}. These figures refer to
the mesh example in Figure \ref{figtest}-(c). The general energy
minimization progress is shown in Figure \ref{vc-conv}. In this
example the driving routine alternates between two grid size
V-cycles: each odd V-cycle solves the correction problem for the
16x16 grid, while even V-cycles improve the previous iterations
with the grid 32x32.
Figures \ref{wr-conv} and \ref{wrb-conv} show the
energy behavior of the Window relaxations (without V-cycles) for
16x16 grid iterations and alternately 16x16 and 32x32 grid sizes,
respectively. Clearly, the V-cycle algorithm is more powerful in
minimizing the energy than just employing the Window relaxations.
\par A more complicated example is shown in Figure
\ref{figm64add} in which the 64x64 mesh graph randomly perturbed
by vertex shifts (up to $2h_x$ of a 64x64 grid), compressed at the
left bottom corner and augmented by 50 randomly chosen edges
 (Figure \ref{figm64add}-(a)). The final result of the
 algorithm is presented in Figure  \ref{figm64add}-(b), where
  all vertices are placed almost at their optimal locations (note the different scales of the two figures).
 We have
  used 2-FMG cycles with 2 V-cycles at each level as the main driving routine.
  Such a driving routine works with the following
  grid sizes: 2, 2, 4, 4, 8, 8, ..., 128, 128, 2, 2, 4, 4, and so forth. After these 2-FMG cycles
  the total energy was very close to its real minimum and
  additional iterations have only slightly corrected the layout.  The next experiment consists of the 64x64 compressed
mesh with three holes. The initial and final layouts are
 presented in Figures \ref{figm64hole}-(a) and \ref{figm64hole}-(b), respectively.
\par Two additional examples demonstrate the layout corrections
for graphs whose vertices have nonequal volumes (see Figures \ref{figvoltest}
and \ref{figvoltest-tree}). In both cases the initial layout
of these graphs was random.
\par In spite of the promising results presented above,
the algorithm has not yet been optimized. However, it is already
clear that several parameters must for efficiency be kept very
small. For example: (1) the number of {\it Window relaxation}
iterations should be fixed between 1 and 3; (2) "optimal enough"
in {\bf SingleWindowSolver} means less than 6 iterations and (3)
the size of $\wnd$ in {\bf SingleWindowSolver} is very robust,
that is, the same results can be obtained with sizes $4$x$4$, $8$x$8$,
and $16$x$16$. We have used only $4$x$4$ as it runs the fastest.




\section{Conclusions}\label{future-work}

\par We have presented a linear time multilevel algorithm
for solving correction to the nonlinear minimization problem
under planar (in)equality constraints. By introducing a sequence
of grids over the domain and a new set of global displacement
variables defined at those grid points, we formulated the
minimization problem under planar equidensity constraints and
solved the resulting system of equations by multigrid techniques.
This approach enabled fast collective corrections for the optimization
objective components. We believe that this formulation can open a
new direction for the development of fast algorithms for efficient
space utilization goals. Among many possible motivating
applications
\cite{gd-book,eades1984,harel88,harelinger,dreznerfacility,meguerdichian01coverage,
cardei-energyefficient, citeulike:717044,vlsi2007book} we focused
on the demonstration of the method on the graph visualization
problem with efficient space utilization demand.
\par We recommend this multilevel method as a general practical
tool in solving, possibly together with other tools, the
nonlinear optimization problem under planar (in)equality
constraints.

\section{Acknowledgments}
\par This work was supported in part by the Office of Advanced Scientific Computing Research, Office of Science, U.S. Department of Energy, under Contract DE-AC02-06CH11357.
\bibliography{mybib}
\hspace*{1.5in}{\scriptsize\framebox{\parbox{2.4in}{
The submitted manuscript has been created in part by UChicago Argonne, LLC, Operator of Argonne National Laboratory ("Argonne").  Argonne, a U.S. Department of Energy Office of Science laboratory, is operated under Contract No. DE-AC02-06CH11357.  The U.S. Government retains for itself, and others acting on its behalf, a paid-up nonexclusive, irrevocable worldwide license in said article to reproduce, prepare derivative works, distribute copies to the public, and perform publicly and display publicly, by or on behalf of the Government.
}}}

\end{document}